\documentclass[sigconf]{acmart}

\usepackage{caption}
\usepackage{amsmath}
\usepackage[linesnumbered,ruled,vlined]{algorithm2e}
\usepackage{multirow}
\usepackage{graphicx}
\usepackage{xcolor}
\usepackage{colortbl}
\usepackage{tcolorbox}
\usepackage{arydshln} 
\theoremstyle{plain}  

\usepackage{enumitem} 

\usepackage{graphicx} 
\usepackage{multirow}
\usepackage[table,xcdraw]{xcolor}
\usepackage[normalem]{ulem}
\useunder{\uline}{\ul}{}
\theoremstyle{plain}  

\usepackage{booktabs,tabularx}

\newtheorem{theorem}{Theorem}
\newtheorem{remark}{Remark}
\theoremstyle{plain}  

\usepackage{enumitem} 

\usepackage{lineno}

\usepackage{colortbl}
\usepackage{tcolorbox}
\usepackage{balance}
\usepackage{textcomp}

\AtBeginDocument{%
  }

\copyrightyear{2026}
\acmYear{2026}
\setcopyright{cc}
\setcctype{by}
\acmConference[KDD '26]{Proceedings of the 32nd ACM SIGKDD Conference on Knowledge Discovery and Data Mining V.2}{August 09--13, 2026}{Jeju Island, Republic of Korea}
\acmBooktitle{Proceedings of the 32nd ACM SIGKDD Conference on Knowledge Discovery and Data Mining V.2 (KDD '26), August 09--13, 2026, Jeju Island, Republic of Korea}
\acmDOI{10.1145/3770855.3817732}
\acmISBN{979-8-4007-2259-2/2026/08}

\begin{document}

\title[Implicit Fine-tuning via Context Engineering: A Curriculum Learning Framework for MMEA]{Implicit Fine-tuning via Context Engineering: A Curriculum Learning Framework for Multimodal Entity Alignment}


\author{Yunpeng Hong}
\email{hongyp@mail.hfut.edu.cn}
\orcid{0009-0000-6150-418X}
\affiliation{%
  \institution{Key Laboratory of Knowledge Engineering with Big Data (the Ministry of Education of China), Hefei University of Technology}
  \city{Hefei}
  \state{Anhui}
  \country{China}
}

\author{Chenyang Bu}
\email{chenyangbu@hfut.edu.cn}
\orcid{0000-0001-8203-0956}
\authornotemark[1]
\affiliation{%
  \institution{Key Laboratory of Knowledge Engineering with Big Data (the Ministry of Education of China), Hefei University of Technology}
  \city{Hefei}
  \state{Anhui}
  \country{China}
}

\author{Di Wu}
\email{wudi.cigit@gmail.com}
\orcid{0000-0002-7788-9202}
\affiliation{%
  \institution{College of Computer and Information Science, Southwest University}
  \city{Chongqing}
  \country{China}
}

\author{Yi He}
\email{yihe@wm.edu}
\orcid{0000-0002-5357-6623}
\affiliation{%
  \institution{Department of Data Science, College of William and Mary}
  \city{Williamsburg}
  \state{VA}
  \country{USA}
}
\author{Xindong Wu}
\authornote{Corresponding authors.}
\email{xwu@hfut.edu.cn}
\orcid{/0000-0003-2396-1704}
\affiliation{%
  \institution{Key Laboratory of Knowledge Engineering with Big Data (the Ministry of Education of China), Hefei University of Technology}
  \city{Hefei}
  \state{Anhui}
  \country{China}
}

\renewcommand{\shortauthors}{Yunpeng Hong et al.}


\begin{abstract}
\begin{sloppypar}
Multimodal Entity Alignment (MMEA) aims to identify equivalent entities across different modalities. While existing methods enhance MMEA performance through black-box context engineering strategies, their reliance on LLM parameter capacity and lack of theoretical interpretability remain unresolved. To this end, we first theoretically validate the mathematical equivalence between context engineering and model fine-tuning in MMEA tasks, demonstrating that prompt components simulate contrastive learning-based sequential fine-tuning in MMEA. Building on this foundation, we then propose PTFEA, a curriculum-learning-inspired framework that translates fine-tuning strategies into interpretable context engineering. Specifically, adaptive difficulty modulation dynamically adjusts information injection stages using confidence thresholds, establishing mathematical equivalence between curriculum learning weights and context sample selection; and three-stage progressive inference incorporates entity information from simple to complex cases, mirroring the gradient descent process in fine-tuning. Experiments on \textbf{five} public datasets demonstrate that PTFEA consistently outperforms strong baselines. In particular, on the ICWIKI dataset, PTFEA narrows the $H@1$ gap between Qwen2.5-72B and 14B to \textbf{0.6\%}. Moreover, compared with the representative context-engineering-based MMEA method MM-ChatAlign, PTFEA reduces the runtime of Qwen2.5-72B \textbf{from 21 hours to 1 hour} and lowers token consumption from 2200–3000 to 200–400, achieving over 80\% reduction on the ICWIKI dataset. This work provides the first theoretical framework unifying context engineering and fine-tuning in MMEA, paving the way for future research that seeks to translate additional fine-tuning strategies into context engineering paradigms. Our code is available at \url{https://github.com/DMiC-Lab-HFUT/PTFEA}.
\end{sloppypar}
\end{abstract}






\maketitle
\begin{sloppypar}
\section{Introduction}
Multimodal entity alignment (MMEA) aims to identify equivalent entities across modalities, playing a crucial role in integrating heterogeneous multimodal data for knowledge-driven applications~\cite{ektefaie2023multimodal,retrieval,zhuo2026knowledge,DBLP:conf/www/ZhuoWWP025,DBLP:conf/ijcai/ZhuoPWW0LW025,chen2026dual,zang2026medical}. To balance efficiency and accuracy, recent work~\cite{jiang2024mm} has begun to explore a hybrid paradigm that combines embedding-based retrieval with large language model (LLM)-based re-ranking. Under this paradigm, multimodal embeddings are first used to retrieve a coarse candidate set, after which LLMs perform fine-grained alignment by reasoning over structured prompts. This two-stage design makes it feasible to incorporate LLMs into MMEA under realistic efficiency constraints.

\begin{figure}
    \centering
\includegraphics[width=\linewidth]{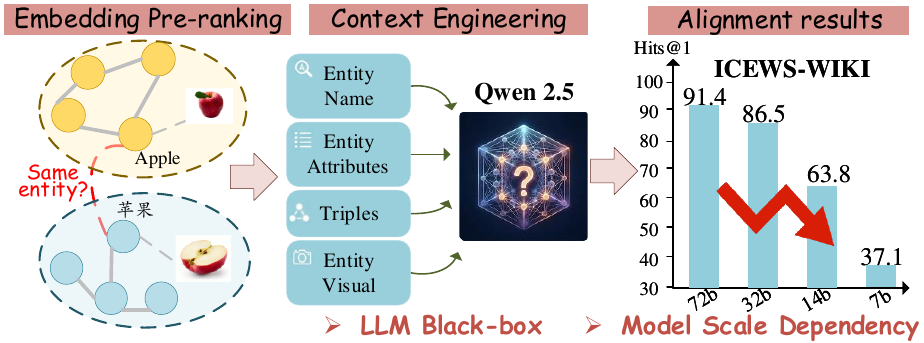}
    \vspace{-2em}
    \caption{Example illustrating that the black-box nature of LLMs makes it difficult to select effective CE strategies, and poor selection can yield substantial performance gaps across model scales using the same prompt.
    }
    \label{fig:motivate}
    \vspace{-2em}
\end{figure}

The re-ranking stage in this paradigm~
\cite{llmea,jiang2024toward,chen2024llm,jiang2024unlocking} typically relies on context engineering (CE)~\cite{DBLP:conf/aaai/HuangBHZW26,DBLP:conf/ijcai/HuangB0025}, where prompts incorporate entity-related information (such as names, local graph structure, and multimodal attributes) to guide alignment decisions. 
For instance, MM-ChatAlign~\cite{jiang2024mm} concatenates heterogeneous evidence associated with each candidate entity into a fixed-order prompt and asks an LLM to determine the match. However, compared with the growing body of work in single-modality entity alignment~\cite{NEURIPS2024_1b57aadd,jiang2024unlocking,yang2024advancing}, studies adopting LLM-based re-ranking in the multimodal setting remain relatively limited. A key reason lies in the high cost of processing multimodal inputs, especially images. In addition to efficiency concerns, existing CE strategies in MMEA are largely heuristic and lack theoretical grounding. When multimodal evidence is rich and heterogeneous, naively injecting multiple sources of information into prompts often introduces noise and redundancy, making alignment performance strongly dependent on LLM parameter scale. As illustrated in Fig.~\ref{fig:motivate}, under the same CE setup, Qwen-2.5 shows a \textbf{54.3\%} $H@1$ gap between the 72B and 7B variants. 

Recent studies in the CE literature have shown that, under certain assumptions, prompting can be theoretically interpreted as being equivalent to model fine-tuning~\cite{ren2024towards,dai2022can}. However, whether this equivalence holds in the MMEA setting remains unexplored. More importantly, even if such an equivalence can be established, how to leverage it to design CE strategies inspired by principled fine-tuning techniques remains an open problem. Addressing this gap is essential for moving beyond heuristic prompt design and toward theoretically grounded and capacity-robust CE methods for MMEA.



In this paper, we first establish the theoretical framework that unifies CE and model fine-tuning in the MMEA setting, achieved by extending the proof technique from \cite{ren2024towards} to MMEA. We theoretically demonstrate that, when processing heterogeneous entity information, the softmax attention in LLMs is mathematically equivalent to the linear attention layer of a fine-tuned model. Under this equivalence, each prompt component in MMEA can be interpreted as performing an implicit fine-tuning step, and a multi-component prompt corresponds to a sequence of such updates.

Once the equivalence for sequential components is
established, the ordering of prompt components becomes a critical design factor, particularly for capacity-limited LLMs. This is because MMEA naturally exhibits evidence with varying difficulty and discriminative power. If harder or noisier evidence is introduced prematurely, the implicit fine-tuning process induced by prompting may be dominated by high-variance and potentially conflicting updates, leading to unstable behavior and degraded alignment performance. In contrast, a well-established principle in model fine-tuning is to organize training signals in an easy-to-hard manner, so as to stabilize optimization and guide the model along a more favorable learning trajectory ~\cite{sahoo2025the,agarwal2022estimating}. From this perspective, structuring prompt components from core, high-confidence information to increasingly complex contextual evidence directly instantiates curriculum learning~\cite{xia2025improving, Curriculum_learning,wang2025end} within CE. This observation provides a principled explanation for why curriculum learning is particularly well suited to MMEA, and clarifies how fine-tuning strategies can be systematically translated into effective and interpretable CE designs.

Based on this insight, we propose PTFEA, a curriculum-learning CE framework that explicitly translates a fine-tuning strategy into prompt construction to alleviate performance gaps caused by LLM parameter scale~\cite{guo2020fine,nair-etal-2024-curriculum}. 
PTFEA allows the LLM, especially smaller ones, to first attempt alignment with minimal, high-confidence
information (Stage I) and progressively introduces more complex evidence when necessary (Stages II \& III). This design mirrors the ``multi-round incremental information injection” paradigm observed in human–LLM interaction.
Experimental results validate our theoretical analysis, showing that PTFEA achieves state-of-the-art performance on 5 public datasets, outperforming strong baselines.  On ICWIKI, the H@1 gap between Qwen2.5-72B and 14B under PTFEA is only \textbf{0.6\%}. Meanwhile, compared with MM-ChatAlign~\cite{jiang2024mm}, PTFEA reduces the runtime of Qwen2.5-72B from 21 hours to 1 hour and cuts token consumption from the 2200–3000 range to 200–400, achieving over 80\% reduction. 
Ablation results further confirm the effectiveness of each component and the practical benefits of our theoretically grounded design.


\noindent
\textbf{Specific contributions} of this paper are as follows.
\begin{itemize}[left=0pt, label=•,topsep=0pt, itemsep=0.1em]
     \item We formally establish the mathematical equivalence between context engineering and model fine-tuning in MMEA tasks. To the best of our knowledge, this provides the first theoretical unification of these two paradigms in this domain. Our analysis lays a foundation for future research to design more effective context engineering strategies by leveraging insights from fine-tuning.
    \item  From the perspective of a representative fine-tuning strategy, curriculum learning, we design a context engineering strategy for MMEA tasks, thereby empirically validating our theoretical equivalence established in our framework.
    \item Experimental results demonstrate that PTFEA achieves state-of-the-art performance on five public datasets, substantiating the practical value of unifying fine-tuning and context engineering in LLM-driven MMEA tasks.
\end{itemize}


\section{Related Work}
MMEA~\cite{chen2020mmea,ack-mmea,AttrMMEA,zhao2025me3a,zhang2025graph,chen2025noise,hong2026psqe} aims to identify and align equivalent entities across different knowledge graphs \cite{DBLP:conf/acl/BuCCDW0025} by jointly leveraging information from multiple modalities, such as textual descriptions and visual representations.
Existing MMEA methods can be broadly categorized into two paradigms: knowledge representation learning-based MMEA (KRL-based MMEA) and LLM re-ranking based MMEA (LLM-based MMEA).

\subsection{KRL-based MMEA}
KRL-based MMEA methods~\cite{su2024ibmea,LoginMEA,pmf} aim to learn entity representations in a multimodal semantic space and perform cross knowledge graph entity matching based on representation similarity.
One mainstream line of research focuses on modeling the structural information of knowledge graphs and fusing textual, visual, and other multimodal features. Representative methods include EVA~\cite{liu2021visual}, MCLEA~\cite{mclea}, MEAformer~\cite{chen2023meaformer}, PSNEA~\cite{ni2023psnea}, SGMEA~\cite{cheng2025sgmea}, CateEA~\cite{feng2025cateea}, and IBMEA~\cite{su2024ibmea}, which typically employ graph neural networks to encode graph structures and jointly learn unified entity representations from multimodal information.
Another line of work places greater emphasis on textual semantic modeling or adopts alternative structural encoding strategies. MMIEA~\cite{zhu2023mmiea} leverages BERT-based encoders to model textual features such as entity names, attributes, and relations, and further refines entity representations through adjacency matrix-based interactions. Simple-HHEA~\cite{jiang2024toward} adopts a biased random walk strategy to obtain graph structural embeddings, capturing more complex structural relationships among entities.
In addition, LGEA~\cite{lu2025breaking} and TMEA~\cite{TMEA} introduce LLMs to infer attribute alignment information and supplement visual information, thereby further enhancing multimodal entity representations.
With the rapid advancement of LLMs, leveraging their strong semantic understanding capabilities for entity alignment has become increasingly attractive.

\vspace{-1em}
\subsection{LLM-based MMEA}
Due to the large scale of knowledge graphs and the limited context window of LLMs, LLM-based MMEA methods typically follow a two-stage pipeline: candidate entity pairs are first generated by knowledge representation learning and then reranked by LLMs leveraging their language understanding and background knowledge.
Current approaches of this kind are still under active development, primarily focusing on LLM-based unimodal entity alignment methods.
For example, Seg-Align~\cite{yang2024advancing} reformulates the entity alignment task as a multiple-choice question, prompting LLMs to select the most appropriate aligned entity from a set of candidates. ChatEA~\cite{jiang2024unlocking} leverages alignment examples to stimulate multi-step reasoning within LLMs based on in-context information. LLMAlign~\cite{chen2024llm} incorporates structured information such as entity attributes and relations to guide the model through multi-round voting, thereby identifying likely aligned entity pairs and mitigating the hallucination effects observed in LLMs.

To the best of our knowledge, research on LLM-based MMEA remains relatively limited, with examples such as MM-ChatAlign~\cite{jiang2024mm}, which adopts CLIP~\cite{radford2021learning} for multimodal similarity modeling and uses GPT-4V~\cite{yang2023dawn} to verbalize visual information, improving LLM-based alignment discrimination.
The limited progress in LLM-based MMEA may stem from the rich and heterogeneous nature of multimodal evidence—which makes prompt design fragile—and the difficulty for smaller LLMs to effectively understand such information under constrained context budgets.

\section{Theoretical Analysis}
\label{Theoretical Analysis}

\subsection{Preliminary}

\textbf{Problem Formulation.}
We define a $MMKG$ as $KG = (E, R, A, V)$, where each ${\rm{e}} \in E$, ${\rm{r}} \in R$, ${\rm{a}} \in A$, and ${\rm{v}} \in V$ denotes an entity, a relation, an attribute, and a visual respectively.
Given two multi-modal knowledge graphs $KG_s = (E_s, R_s, A_s, V_s)$ and $KG_t = (E_t, R_t, A_t, V_t)$, the task of multimodal entity alignment targets to discover the set of equivalent entity pairs between $KG_s$ and $KG_t$, denoted as $\mathcal{M}=\{ (e_s^i,e_t^j)| e_s^i \equiv e_t^j, e_s^i \in E_s, e_t^j \in E_t\}$, where $e_s^i \equiv e_t^j$ means an equivalent entity between $e_s^i$ and $e_t^j$.

\textbf{CE of LLMs.}
Given an input sequence $X$ containing specific information about the candidate entities, the goal is to generate the final reordered alignment result $H$. The input sequence for MMEA can be represented as:
\begin{equation}
    X = [X_{1},..., X_{k}, X_T]
\end{equation}
where $X_i = [x_i^1, x_i^2, \ldots, x_i^n]$ denotes a subsequence consisting of $n$ descriptive tokens, representing prompt information such as task descriptions or output constraints, and $X_T = [x_T^1, x_T^2, \ldots, x_T^m, x_{T}^{m+1}]$ denotes a subsequence containing $m+1$ query tokens. This input sequence is fed into an LLM to generate the final alignment output $H = LLM(X)$.
Since LLMs are typically built upon multi-layer Transformer architectures, we simplify the representation by considering a softmax attention layer~\cite{ren2024towards}. The computation can be formalized as:
\begin{equation}
    H = \text{Transformer}(X) = W_V X \cdot \text{softmax} \left( \frac{(W_K X)^\top W_Q X}{\sqrt{d}} \right)
\end{equation}
where $W_K, W_Q, W_V$ are the projection matrix.
For the query input at position $m+1$, we extract the corresponding query vector $x_{T}^{m+1}$ from $X_T$ to generate the prediction label, since in practical settings, this output representation is often aligned with next-token prediction tasks~\cite{ren2024towards}. Specifically, the final output vector ${\boldsymbol{h}}_{T}^{m+1}$ can be formulated as:
\begin{equation}
\label{final_equ_h}
    {\boldsymbol{h}}_{T}^{m+1} = \boldsymbol{W}_{V} \boldsymbol{X} \cdot \text{softmax} \left( \frac{(\boldsymbol{W}_{K} \boldsymbol{X})^{\top} \boldsymbol{W}_{Q} x_T^{m+1}}{\sqrt{d}} \right)
\end{equation}

\textbf{In-context Learning as Fine-Tuning.}
Recent studies~\cite{ren2024towards,dai2022can} establish a duality between gradient descent on a softmax attention linear layer and linear attention fine-tuning. The single-layer softmax attention linear model is defined as~\cite{ren2024towards}:
\begin{equation}
    f(x) = W \phi(x)
\end{equation}
where $\phi(x)=e^{\boldsymbol{w}^{T}\boldsymbol{x} - \|\boldsymbol{x}\|^{2}/2}$ is a specific softmax kernel mapping function that satisfies $exp(x^Ty)= \phi(x)^T \phi(y)$.
Therefore, consistent with~\cite{dai2022can,ren2024towards}, the output of the softmax attention can be rewritten:
\begin{equation}
\small
W_V X \cdot \mathrm{softmax}\left( \dfrac{(W_K X)^\top W_Q X}{\sqrt{d}} \right) 
  \;\Leftrightarrow\;
  LA(c W_V X,\phi({W_K X})^\top,\phi(W_Q X))
\end{equation}
where \( LA(V, K, Q) = \left( \sum\limits_{j = 1}^N v_j \otimes k_j \right) Q \) denotes linear attention, and the normalization factor is given by \( c = \left( \mathbf{1}_N^\top \phi({W_K X})^\top \phi(W_Q X) \right)^{-1} \).

Moreover,~\cite{ren2024towards} elaborates on the loss function form corresponding to the fine-tuning model.
 It shows that, given a demonstration set \( X_D = [x^i]_{i=1}^N \) and a query set \( X_T \), the query token $x_{T}^{m+1}$ passing through a simplified LLM structure consisting of a single softmax attention layer is equivalent to one gradient descent step on the dual model $ f(x) = W \phi(x)$. The corresponding loss function is formulated as $\mathcal{L} = - \frac{1}{\xi C} \sum_{i=1}^N  (W_V x^i)^T  W \, \phi(W_K x^i)$,
where \(\xi\) denotes the learning rate, and \(C\) is a constant.

However, in contrast to \cite{ren2024towards}'s homogeneous
examples, MMEA requires integrating diverse components from two KGs, including attributes,
structures, and visual information. 
We next introduce Theorem~\ref{theorem1} to formalize this connection, showing that the equivalence holds for our structured, multi-component prompt.

\subsection{Context Engineering as Fine-Tuning in MMEA}
\label{Prompting as Fine-Tuning in MMEA}
In this subsection, we show that in MMEA, a simplified LLM (e.g., a softmax attention) leads to a loss function consistent with fine-tuning under linear attention. 
Building on Theorem~\ref{theorem1} and Remark~\ref{remark2}, we observe an alignment between the resulting fine-tuning dynamics and curriculum learning, and thus develop corresponding CE strategies through a curriculum-driven fine-tuning lens.

\begin{figure*}[!t]
\centering
\includegraphics[width=0.92\linewidth]{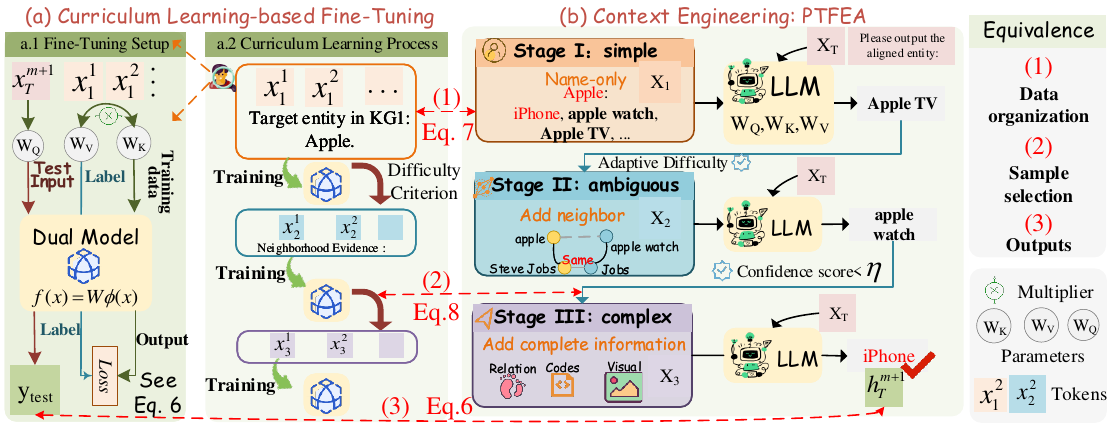}
\vspace{-1em}
\caption{Illustration of the equivalence between curriculum learning-based fine-tuning and the proposed context engineering framework PTFEA. (a.1) the overall training and inference setup of fine-tuning, (a.2) the curriculum-learning-based fine-tuning procedure, and (b) the PTFEA pipeline. The equivalence is reflected in three aspects: (1) consistent data organization, (2) analogous sample selection mechanisms, and (3) aligned outputs.
}
\label{framework}
\vspace{-1em}
\end{figure*}

\begin{theorem}
\label{theorem1}
    Given the entity list information $X= [X_{1},..., X_{k}, X_T]$, the output representation ${\boldsymbol{h}}_{T}^{m+1}$ obtained through a single-layer softmax attention inference is equivalent to the test prediction ${y}_{{test}}$ produced by the dual model $f(x) = W\phi(x)$ after performing a single-step gradient descent under the loss function $\mathcal L$. 
 \begin{equation}
 \small
 \left\{ {\begin{array}{*{20}{c}}
{\boldsymbol{h}_T^{m + 1} \quad \Longleftrightarrow \quad {y_{test}}}\\
{{\mathcal L} =  - \frac{1}{{\xi C}}\sum\limits_{j = 1}^k {{{\mathcal L}_j}}  =  - \frac{1}{{\xi C}}\sum\limits_{j = 1}^k {\sum\limits_{i = 1}^N ( } {W_V}x_j^i{)^ \top }{\bf{W}}{\mkern 1mu} \phi ({W_K}x_j^i)\quad }
\end{array}} \right.
\label{theorem1eq}
 \end{equation}
 where $\xi$ is the learning rate, $\phi(x)$ is a specific softmax kernel mapping function, and $C$ is a constant.
\end{theorem}
The detailed proof can be found in Appendix~\ref{theproof}. The following describes the fine-tuning setup for model equivalence, as shown in Fig.~\ref{framework} (a). Note that during CE, LLM parameters are kept frozen; therefore, we assume that the attention-layer parameters $W_K$, $W_Q$, and $W_V$ are already fixed at initialization.

\subsubsection{Fine-tuning setup}
\label{Fine-tuning setup}
For the setup of the dual model, we focus on three key aspects: the training data (i.e., inputs and labels), the model test input, and the initialization parameters $W_0$.

\textbf{Training data.}
Training samples are derived directly from the CE. Given the context prompt $[X_1,\ldots,X_k]$ with token representations $\{x_j^i\}_{i=1}^{n}$ in $X_j$, we form
$\mathcal{D}_1=\{(W_K x_1^{i},\, W_V x_1^{i})\}_{i=1}^{n}$,
where $W_K x_1^{i}$ and $W_V x_1^{i}$ serve as the \textbf{input} and \textbf{label}, respectively.
This means that the descriptive tokens in our context prompt (e.g., entity names, attributes, or task descriptions) are algorithmically mapped into input-label pairs for the implicit dual model $f(x)$ in our theoretical framework.

The fundamental reason for this setup follows from the form of the loss $L_j$ in Eq.~\ref{theorem1eq}.
In the dual model $f(x)=W\,\phi(x)$, when the input is $W_K x_i$, the model prediction is $W\,\phi(W_K x_i)$, which matches the right-hand side of the inner product in the loss. Therefore, the left-hand side term $(W_V x_i)^{\top}$ naturally serves as the target representation (i.e., the label) for the prediction.
Moreover, this formulation can be interpreted from a \textbf{contrastive learning} perspective. If we view $y_d^i = W\,\phi(W_K x_j^i)$ as the predicted output and $y_l^i = W_V x_j^i$ as the target representation, then the corresponding objective can be written as
$\mathcal{L}_j = \sum_{i=1}^{N} (y_l^i)_j^{\top}(y_d^i)_j$,
i.e., maximizing the similarity between the predicted and target representations.

\textbf{Test input and model initialization parameter $W_0$.}
For the query tokens $X_T=[x_T^1,x_T^2,\ldots,x_T^m,x_T^{m+1}]$. The first $m$ tokens $[x_T^1,x_T^2,\ldots,x_T^m]$ are used to construct the initialization parameter $W_0$ of the dual model. Let $X_T^{(m)}=[x_T^1,x_T^2,\ldots,x_T^m]$, then $W_0 = C'\, W_V X_T^{(m)}\, \phi\!\big(W_K X_T^{(m)}\big)$, where $C'$ is a fixed hyperparameter. The last query token $x_T^{m+1}$ is mainly used as the test input, i.e., we take $W_Q x_T^{m+1}$ as the test input representation.

\begin{remark}
Unlike the loss in~\cite{ren2024towards}, which relies solely on demonstration-case information, the MMEA loss in Theorem~\ref{theorem1} comprises multiple components. As the prompt tokens are fed sequentially into the LLM, these prompt components effectively simulate contrastive learning-based sequential fine-tuning.
\label{remark2}
\end{remark}

Building on Theorem~\ref{theorem1} and Remark~\ref{remark2}, we clarify the prompt organization in MMEA and provide an interpretable view of the otherwise black-box context-engineering process. Leveraging the equivalence between fine-tuning and CE, we can translate fine-tuning principles into context-engineering strategies. 
Our motivation for using curriculum learning comes directly from the structure of MMEA: the evidence has different levels of difficulty and usefulness. This is not a random choice from a set of fine-tuning tricks, but a method that matches the task. (1) Easy-to-hard evidence: entity cues range from simple name matching (easy), to disambiguation using local graph structure (medium), and to combining multimodal attributes for fine-grained decisions (hard). This layered structure naturally supports a curriculum. (2) Reducing the gap across LLM sizes under long inputs: a stage-wise curriculum introduces information gradually, which helps prevent overload and reduces performance differences across model scales.

\subsubsection{CE Form of Curriculum Learning-based Fine-Tuning}
By Theorem~\ref{theorem1}, introducing a control coefficient $v_j\in\{0,1\}$ before $\mathcal L_j$ enables selection and weighting of loss terms and supports an ordered, stage-wise optimization, making the fine-tuning formulation formally consistent with curriculum-style context engineering. Meanwhile, since curriculum-based fine-tuning typically relies on a difficulty criterion for sample selection, we can establish correspondence by introducing a sample selection mechanism in CE.

\begin{remark}
With a difficulty-aware gate $v_j$, curriculum learning and CE admit a direct correspondence in how training signals are scheduled and injected:
\begin{equation}
\small
{\mathcal L}=-\frac{1}{\xi C}\sum_{j=1}^{k} v_j\,{\mathcal L}_j
\quad \Longleftrightarrow \quad
X=\sum_{j=1}^{k} v_j\,X_j
\label{reamrkeq1}
\end{equation}
where curriculum learning uses $v_j$ to select/reweight the $j$-th loss term, while CE uses the same $v_j$ to schedule whether the $j$-th evidence block is composed into the prompt. Moreover, $v_j$ can be determined by a confidence-based difficulty criterion:
\begin{equation}
\small
v_j=\mathbb{I}\!\left(c_j^{(t)}\ge \tau^{(t)}\right)
\quad \Longleftrightarrow \quad
v_j=\mathbb{I}\!\left(c_j^{(t)}\ge \eta\right)
\label{remarkeq2}
\end{equation}
where $c_j^{(t)}$ denotes the model confidence for the $j$-th training signal at stage $t$, $\tau^{(t)}$ is a stage-dependent threshold in fine-tuning, and $\eta$ is a gating threshold in CE. This shows that both curriculum learning and CE implement a sample selection mechanism.
\label{remarkFT}
\end{remark}
Therefore, based on Theorem~\ref{theorem1} and Remark~\ref{remarkFT}, the correspondence between curriculum-based fine-tuning and CE requires:
(i) both to follow the formulation in Eq.~\ref{theorem1eq},~\ref{reamrkeq1} and \textbf{ensuring consistent data organization and outputs};
(ii) both to follow the formulation in Eq.~\ref{remarkeq2}, CE to \textbf{incorporate a compatible sample selection mechanism}, matching the selection process in curriculum-based fine-tuning, as shown in Fig.~\ref{framework}.

\section{Method}
\label{methods}
\subsection{Overview}
Motivated by the theoretical analysis in Sec.~\ref{Theoretical Analysis}, we propose a \textbf{curriculum-driven context engineering framework} that reduces the sensitivity of alignment performance to LLM parameter scales, as shown in Fig.~\ref{framework}. Concretely, we obtain candidate entity sets using MEAformer~\cite{chen2023meaformer} and Simple-HHEA~\cite{jiang2024toward}, and construct an entity-specific alignment candidate pool via a hybrid retrieval strategy (Sec.~\ref{Candidate Entity List Generation}). Inspired by~\cite{cuadron2025danger,wang2019dynamic}, we observe substantial heterogeneity in the amount of contextual evidence required for reliable alignment across entities. Accordingly, we categorize entities into \textit{simple}, \textit{ambiguous}, and \textit{complex} types, and design a three-stage \textbf{progressive evidence augmentation} pipeline (Sec.~\ref{Progressive LLMs Inference}) that increases available information from coarse to fine and from easy to hard. \textbf{This design mirrors curriculum-based fine-tuning}: organizing instances by ascending difficulty enables a gradual strengthening of model capability. Moreover, we introduce an Adaptive Difficulty Modulation mechanism (Sec.~\ref{Knowledge-Driven Alignment Prediction}) to dynamically decide whether additional evidence should be injected, which effectively serves as the difficulty criterion in curriculum learning and yields robust alignment under constrained model capacity.

\vspace{-0.5em}
\subsection{Candidate Entity List Generation}
\label{Candidate Entity List Generation}
In MMEA, inputting all entities at once may exceed the LLM's context window. To tackle this, we follow prior work~\cite{jiang2024unlocking,yang2024advancing,jiang2024mm} by using KRL-based embeddings (e.g., Simple-HHEA~\cite{jiang2024toward}) to pre-filter candidate entities. To improve recall and adapt to diverse data, we further adopt a hybrid retrieval strategy that integrates multiple similarity measures, enhancing overall robustness.

Specifically, given a source entity set $E_s = \{e_s^1, \dots, e_s^m\}$ and a target entity set $E_t = \{e_t^1, \dots, e_t^n\}$, we define a hybrid similarity function to compute the pairwise similarity between each source entity $e_s^i$ and target entity $e_t^j$ as follows:
\begin{equation}
\mathrm{Sim}_{ij} = (1 - \alpha) \cdot \mathrm{KRL^1}_{ij} + \alpha \cdot \mathrm{KRL^2}_{ij},
\end{equation}
where $\mathrm{KRL^1}$ and $\mathrm{KRL^2}$ denote the similarity scores derived from MEAformer~\cite{chen2023meaformer} and Simple-HHEA~\cite{jiang2024toward}, respectively, and $\alpha \in [0, 1]$ is a tunable weighting hyperparameter controlling their relative contributions.
For each source entity $e_s^i \in E_s$, we retrieve the top-$k$ target entities from $E_t$ based on hybrid similarity scores $\mathrm{Sim}_{ij}$. The candidate set $E^i$ is $E^i = \left( e_t^{j_1}, e_t^{j_2}, \dots, e_t^{j_k} \right)$,
where the indices $j_1, j_2, \dots, j_k$ correspond to the $k$ highest similarity scores, ordered as $\mathrm{Sim}_{ij_1} \geq \mathrm{Sim}_{ij_2} \geq \dots \geq \mathrm{Sim}_{ij_k}$. The final retrieval output is the set of candidate lists for all source entities: $\mathcal{E} = \{E^1, E^2, \dots, E^m\}$.

\vspace{-0.5em}
\subsection{Curriculum-Guided Progressive Inference}
\label{Progressive LLMs Inference}
Similar to curriculum learning, we progressively provide information to help LLMs with varying parameter scales gradually grasp semantic content.

\subsubsection{Stage \uppercase\expandafter{\romannumeral 1}: Simple Entity Selection}
\label{Knowledge-Driven Alignment Prediction}
In Stage~\uppercase\expandafter{\romannumeral 1}, we leverage the inherent knowledge of the LLM to select the most probable aligned entity $e_t^j$ from the candidate set $E^j$, along with a confidence score $\mathcal{C}$. If $\mathcal{C}$ is sufficiently high, the prediction is accepted as the final alignment result. This stage effectively identifies ``simple entities'', especially when using LLMs with large parameter sizes, which often require no additional information (e.g., attributes), which reduces token consumption and allows for efficient pre-alignment.

Specifically, we divide the input information to LLM into four parts: Entity Description, Task Specification, Output Format Description, and Task. 
The entity description (ED) includes the source entity $e_s^i$ and its corresponding list of candidate entity names retrieved from $KG_t$. The task specification (TS) instructs the model to select an entity strictly from the candidate list, relying on its background knowledge without generating arbitrary outputs. The output format description (OFD) requires the model to return the selected entity’s name along with a confidence score. The task (T) clarifies the specific action to be performed.
This process can be formalized as:
\begin{equation}
(e, c) = \mathrm{LLM}(ED, TS, OFD, T),
\end{equation}
where $e$ denotes the selected entity name and $c$ represents the associated confidence score.

\textbf{Adaptive Difficulty Modulation.}
Since different LLMs may express varying levels of confidence in their outputs, using a fixed threshold can lead to the following issues: if the threshold is set too high, models that are generally less confident may have their correct answers mistakenly rejected, triggering a stage \uppercase\expandafter{\romannumeral 2} process and resulting in unnecessary resource consumption. Conversely, if the threshold is too low, the model may mistakenly accept incorrect answers as reliable, compromising the final alignment accuracy.

Specifically, we employ an adaptive difficulty modulation strategy by dynamically setting the confidence threshold $\eta_i$ based on the average confidence scores of the last ten predictions. If the average confidence $\mathrm{avg}(C_{i-10}, \ldots, C_{i-1})$ exceeds half of the maximum score $Q$, it is used as the current threshold; otherwise, $\eta_i$ is set to $\frac{Q}{2}$. Formally,
\[
\eta_i =
\begin{cases}
\mathrm{avg}(C_{i-10}, \ldots, C_{i-1}), & \text{if } \mathrm{avg}(C_{i-10}, \ldots, C_{i-1}) > \frac{Q}{2} \\
\frac{Q}{2}, & \text{otherwise}
\end{cases}
\]
If the confidence score $C_i$ for entity $e_s$ falls below $\eta_i$, the entity is considered too difficult to align at the current stage and is deferred to the next stage. Accordingly, the sample selection variable $v_{II}$ is determined by the indicator function:
\[
v_{II} = \mathbb{I}(c^{(I)} < \eta)
\]
which controls whether the sample is activated for training based on its predicted confidence.

\subsubsection{Stage \uppercase\expandafter{\romannumeral 2}: Ambiguous Entity Selection}
\label{Structure-Aware Alignment Prediction}
In scenarios where the model exhibits low confidence during the stage \uppercase\expandafter{\romannumeral 1}, namely when the LLM relies solely on existing entity name information and fails to produce reliable alignment predictions, we consider these as ``ambiguous entities''. It becomes imperative to incorporate structural information to enhance the model’s comprehension of the alignment task. To our knowledge, existing LLM-based approaches for entity alignment have not fully leveraged the training data employed in entity embedding methods. To address this limitation, in the stage \uppercase\expandafter{\romannumeral 2}, we introduce training data associated with the candidate and target entities into LLM. This design serves two primary purposes: (1) to enable the model to \textbf{learn prototypical alignment patterns} from labeled examples and thereby develop a more nuanced understanding of aligned entity pairs; and (2) to \textbf{provide additional structural context} by including information about neighboring entities. The underlying hypothesis is that if two entities share aligned neighbors, they are more likely to be aligned themselves. 
This process can be formalized as:
\begin{equation}
(e, c) = \mathrm{LLM}(ES, ED, TS, OFD, T),
\end{equation}
where $\mathrm{ES}$ denotes the alignment information derived from the local neighborhood structures of the entities. The formal expression is as follows:
\begin{equation}
\mathrm{ES} = \{(e_s^i, e_t^j) \in \mathcal{N}(e_s) \times \mathcal{N}(e_t) \mid (e_s^i, e_t^j) \in \mathcal{A}\}
\end{equation}
where $\mathcal{N}(e)$ denotes the set of neighboring entities of entity $e$, and $\mathcal{A}$ denotes the set of aligned entity pairs provided in the training data.
If LLM's confidence score is greater than or equal to the confidence threshold from Stage \uppercase\expandafter{\romannumeral 1}, we return its prediction result; otherwise, we proceed to Stage \uppercase\expandafter{\romannumeral 3}.

\vspace{-0.5em}
\subsubsection{Stage \uppercase\expandafter{\romannumeral 3}: Complex Entity Selection}
\label{Full-Information Reasoning Alignment Prediction}
In Stage~III, we leverage LLMs to deeply extract and reason over entity evidence and introduce a multi-dimensional, step-wise scoring scheme to comprehensively assess candidates from multiple aspects, including name cues, visual similarity (pre-scored by a lightweight multimodal encoder to reduce LLM token costs), neighborhood context, and KG triples. To further improve robustness, we instantiate an explicit think-then-rethink procedure~\cite{jiang2024unlocking,jiang2024mm}: the model first produces a coarse alignment probability grounded in multimodal semantics and then enters a reflection stage to assess confidence and refine predictions via an iterative candidate-set expansion strategy, yielding a more reliable final decision. For an entity pair $(e_s, e_t)$, we compute its alignment score as:
\begin{equation} \left\{ {\begin{array}{*{20}{c}} {{\rm{Scor}}{{\rm{e}}_i}({e_s},{e_t}) = LLM(ED,Image,relation,...)}\\ {{\rm{Score}}({e_s},{e_t}) = \sum\limits_{i = 1}^k {{w_i}} \cdot {\rm{Scor}}{{\rm{e}}_i}({e_s},{e_t})} \end{array}} \right. 
\label{socre12}
\end{equation}
where $\text{Score}_i(e_s, e_t)$ denotes the modality-specific score from the $i$-th perspective (e.g., name similarity, image similarity, and relation matching), inferred by the LLM conditioned on the provided entity evidence, and $w_i$ is the corresponding weight.

\section{Experiments}

\subsection{Experimental Setup}
\subsubsection{Datasets}

We conducted experiments on five popular benchmark datasets, including three subsets from DBP15K~\cite{liu2021visual}: French-English (FR-EN), Japanese-English (JA-EN), and Chinese-English (ZH-EN), as well as two additional datasets: ICEWS-WIKI (ICWIKI) and ICEWS-YAGO (ICYAGO)~\cite{jiang2024toward}.
DBP15K is a widely used benchmark dataset for cross-lingual multimodal entity alignment tasks. Although it does not contain raw image data, images have been preprocessed into vector representations and stored accordingly. In contrast, ICWIKI and ICYAGO exhibit significant differences in structural modality and other modal features, and both include raw image data.
Detailed statistics of each dataset can be found in Appendix~\ref{dataset}.


\vspace{-0.5em}
\subsubsection{Baselines}
\label{baselinesH}
We compare our approach against 20 baseline methods. On the DBP15K dataset, we evaluate our model alongside both embedding-based and LLM-based methods. Specifically, BootEA~\cite{sun2018bootstrapping}, SDEA~\cite{sdea}, and SCMEA~\cite{zhou2024scmea} are unimodal embedding-based approaches, while EVA~\cite{liu2021visual}, MSNEA~\cite{MSNEA}, MCLEA~\cite{mclea}, MEAformer~\cite{chen2023meaformer}, DESAlign~\cite{wang2024towards}, CDMEA~\cite{su2025mitigating}, OTMEA~\cite{wang2025otmea}, and $AMF^2SEA$~\cite{li2025exploring} are multimodal embedding-based methods. In addition, LLMEA~\cite{llmea} and Seg-Align~\cite{yang2024advancing} represent LLM-based entity alignment approaches.
For the ICWIKI and ICYAGO datasets, the selected embedding-based baselines include Dual-AMN~\cite{mao2021boosting}, TEA~\cite{TEA}, XGEA~\cite{xgea}, MMIEA~\cite{zhu2023mmiea}, MEAformer, and Simple-HHEA~\cite{jiang2024toward}. The LLM-based baselines include ChatEA~\cite{jiang2024unlocking} and MM-ChatAlign~\cite{jiang2024mm}.
We adopt the most commonly used evaluation metrics in the MMEA literature, namely Hits@N ($H@N$ $\uparrow$) and Mean Reciprocal Rank ($MRR$ $\uparrow$).

\begin{table}[htbp]
\caption{Comparison of PTFEA with 8 baselines on the ICWIKI and ICYAGO datasets.}
\vspace{-1em}
\setlength{\tabcolsep}{4pt}
\begin{tabular}{ccccccc}
\hline
\multicolumn{1}{c}{} & \multicolumn{3}{c}{ICWIKI} & \multicolumn{3}{c}{ICYAGO} \\
\multicolumn{1}{c}{\multirow{-2}{*}{Models}} & H@1 & H@10 & MRR & H@1 & H@10 & MRR \\ \hline
 Dual-AMN & .083 & .281 & .145 & .031 & .144 & .068 \\
 TEA  & .610 & .894 & .718 & .657 & .891 & .740 \\
 XGEA & .549 & .628 & .575 & .314 & .421 & .351 \\
 MMIEA  & .562 & .716 & .616 & .745 & .857 & .787 \\
 MEAformer & .644 & .842 & .713 & .698 & .878 & .762 \\
 Simple-HHEA & .720 & .872 & .754 & .847 & .915 & .870 \\ \hline
 ChatEA & .880 & .945 & .912 & .935 & .955 & .944 \\
 MM-ChatAlign & {\ul .945} & {\ul .966} & {\ul .948} & {\ul .930} & {\ul .965} & {\ul .943} \\
 \cellcolor[HTML]{C0C0C0}PTFEA & \cellcolor[HTML]{C0C0C0}\textbf{.984} & \cellcolor[HTML]{C0C0C0}\textbf{.993} & \cellcolor[HTML]{C0C0C0}\textbf{.985} & \cellcolor[HTML]{C0C0C0}\textbf{.978} & \cellcolor[HTML]{C0C0C0}\textbf{.985} & \cellcolor[HTML]{C0C0C0}\textbf{.980} \\ \hline
\end{tabular}
\label{icews}
\end{table}

\subsubsection{Implementation Details}
\label{Implementation Details}
LLM inference was conducted through API calls, while the A40 GPU was used for embedding-based retrieval and visual feature computation.
Our main experiments are conducted based on the Qwen2.5-72B. To ensure reproducibility, we fix the random seed to 42 and split each of the five datasets into training and test sets with a ratio of 3:7.
For visual encoding, we use ResNet-152~\cite{He_2016_CVPR} as the visual encoder for the DBP15K dataset. In contrast, for the ICWIKI and ICYAGO datasets, we adopt CLIP~\cite{radford2021learning} to jointly encode entity names and images.
During candidate entity generation, the hyperparameter $\alpha$ is set to 0.0 for the DBP15K dataset and 0.2 for both ICWIKI and ICYAGO. 
The window size is uniformly set to 5.
During re-ranking, the visual information is represented by the similarity between the two images.

\begin{table*}[]
\caption{Performance comparison between PTFEA and 13 baseline methods (Embedding, LLM) on the three DBP15K datasets. $*$ denotes reproduced results, ``MM” indicates multimodal methods, and $\checkmark$ highlights multimodal ones.}
\vspace{-1em}
\centering
\begin{tabular}{cccccccccccc}
\hline
\multicolumn{2}{c}{} &  & \multicolumn{3}{c}{ZH-EN} & \multicolumn{3}{c}{JA-EN} & \multicolumn{3}{c}{FR-EN} \\
\multicolumn{2}{c}{\multirow{-2}{*}{Models}} & \multirow{-2}{*}{MM} & H@1 & H@10 & MRR & H@1 & H@10 & MRR & H@1 & H@10 & MRR \\ \hline
 & BootEA \tiny(IJCAI, 18) & × & .629 & .848 & .703 & .622 & .854 & .701 & .653 & .874 & .731 \\
 & EVA \tiny(AAAI, 21)& \checkmark & .929 & {\ul.983} & .946 & .964 & \textbf{ .997} & {\ul .976} & {\ul .990} & \textbf{.999} & \textbf{.994} \\
 & MSNEA \tiny(KDD, 22)& \checkmark & .887 & .961 & .913 & .938 & .983 & .955 & .969 & .997 & .980 \\
 & MCLEA \tiny(COLING, 22)& \checkmark & .926 & {\ul.983} & .946 & .961 & {\ul .994} & .973 & .987 & \textbf{.999} & .992 \\
  & SDEA \tiny(ICDE, 22) & × & .870 & .966 & .910 & .848 & .952 & .890 & .969 & .995 & .980 \\
 & MEAformer* \tiny(MM, 23)& \checkmark & .931 & \textbf{ .988} &  {\ul.953} &  {\ul.965} & \textbf{ .997} & \textbf{ .978} & .985 & \textbf{.999} & .991 \\
 & DESAlign \tiny(ICDE, 24)& \checkmark & .826 & .972 & .885 & .811 & .963 & .869 & .810 & .987 & .865 \\
\multirow{-8}{*}{\rotatebox{90}{Embedding}} & SCMEA \tiny(Neural Networks, 2024) & × & .908 & .979 & .935 & .934 & .987 & .954 & .982 & \cellcolor[HTML]{FFFFFF}{\ul .998} & .988 \\
& OTMEA \tiny (ICASSP, 25) & \checkmark  & .779 & .953 & .842 & .783 & .959 & .849  & .796 & .970 & .861 \\
& $AMF^2SEA$ \tiny(COLING, 25) & \checkmark  & .691 & .879 & .751 & .696 & .871 & .757  & .767 & .914 & .818 \\
& CDMEA \tiny(SIGIR, 25) & \checkmark & .865 & .978 & .909 & .863 & .983 & .910 & .877 & .989 & .922 \\
\hline
 & LLMEA \tiny(arXiv, 24)& × & .898 & .923 & - & .911 & .946 & - & .957 & .977 & - \\
 & Seg-Align \tiny(EMNLP, 24) & × & {\ul .953} & - & - & .907 & - & - & .987 & - & - \\
\multirow{-3}{*}{\rotatebox{90}{LLM}} & \cellcolor[HTML]{C0C0C0}PTFEA & \cellcolor[HTML]{C0C0C0}\checkmark & \cellcolor[HTML]{C0C0C0}\textbf{.957} & \cellcolor[HTML]{C0C0C0}.966 & \cellcolor[HTML]{C0C0C0}\textbf{ .960} & \cellcolor[HTML]{C0C0C0}\textbf{.977} & \cellcolor[HTML]{C0C0C0}.981 & \cellcolor[HTML]{C0C0C0} \textbf{ .978}  & \cellcolor[HTML]{C0C0C0}\textbf{.992} & \cellcolor[HTML]{C0C0C0}.995 & \cellcolor[HTML]{C0C0C0}{\ul .993} \\ \hline
\end{tabular}
\label{DBP15K_EXPER}
\normalsize
\end{table*}

\subsection{Main Experiment Results}
\begin{tcolorbox} [
    colback=blue!5!white,
    boxsep=0.3pt 
] 
RQ1: How effective is PTFEA on MMEA, and does Theorem~\ref{theorem1} imply an approximate equivalence between CE and fine-tuning?
\end{tcolorbox}
\vspace{-1em}

\subsubsection{Main Results}
To address RQ 1, we conducted experiments on five publicly available datasets and compared our method against 20 algorithms, as shown in Tab.~\ref{icews} and~\ref{DBP15K_EXPER}. The results demonstrate that our approach achieves strong performance compared to the latest MMEA methods.
For example, in Tab.~\ref{icews}, for the multimodal datasets ICWIKI and ICYAGO, PTFEA outperforms the latest multimodal LLM method, MM-ChatAlign, with \textbf{improvements of 3.9\% and 4.8\% in $H@1$}, respectively.
In Tab.~\ref{DBP15K_EXPER}, on the ZH-EN, JA-EN, and FR-EN subsets of the DBP15K dataset, PTFEA achieves \textbf{improvements of 2.6\%, 1.2\%, and 0.2\% in $H@1$}, respectively, compared to the best existing embedding-based methods. PTFEA yields \textbf{improvements of 0.4\%, 7.0\%, and 0.5\%} in $H@1$ over the strongest single-modal LLM-based baseline, Seg-Align, respectively. It is worth noting that PTFEA performs slightly worse on the $H@10$ metric. This is primarily because, in Stage III, even when some complex entities appear within the top-10 candidates during initial retrieval (Eq.~\ref{socre12}), the LLM still struggles to accurately assess their similarity, leading to final rankings that may underperform those based solely on embedding models.
For instance, an entity pair with low name similarity but high relational similarity might be down-ranked if the name weight is predominant. 
This phenomenon indicates that, when dealing with difficult samples, LLMs cannot rely solely on their ranking capability. Nonetheless, this issue does not arise in Tab.~\ref{icews}, mainly because entity names in the ICWIKI and ICYAGO datasets exhibit high discriminability, which reduces ambiguity in the alignment process.
 \textit{These strong results indicate that PTFEA is capable of effectively capturing and aligning entity information across a variety of dataset scenarios.}

\begin{figure}
\centering
\includegraphics[width=0.6\linewidth]{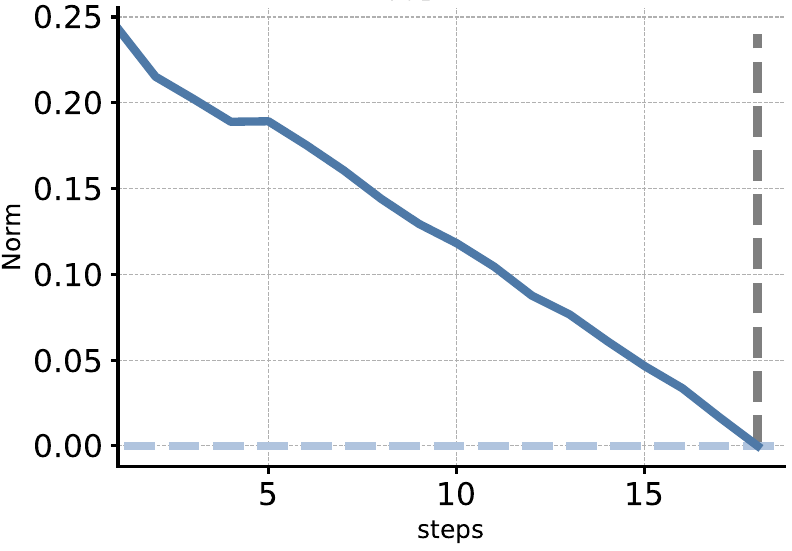}
\vspace{-1.5em}
\caption{
The representation gap between fine-tuning $y_{test}$ and CE ${\boldsymbol{h}}_{T}^{m+1}$ shrinks over training.
}
\label{threoms_ex}
\vspace{-1em}
\end{figure}

\subsubsection{Prompting $\approx$ Fine-tuning}
\label{proisFT}
To validate the theoretical proof presented in Theorem~\ref{theorem1}, we compare the outputs of the actual trained dual model and the softmax attention layer, as illustrated in Fig.~\ref{threoms_ex}. 
We follow the setup of~\cite{ren2024towards} and formulate a linear regression task with a softmax-based self-attention layer. Specifically, we encode $K$ prompt messages using a pretrained FastText model into 100-dimensional vectors. For each entity pair $(e_s^i, e_t^i)$, we compress the name embeddings on both sides to 50 dimensions and concatenate them into a $2\times 50$ (i.e., 100-d) representation, forming an input matrix $\mathbf{X}\in\mathbb{R}^{(K+T+1)\times 100}$, where the first $N=K+T=18$ tokens correspond to the source/target KG inputs and the last token is the query with its label masked (i.e., the last vector is set to zero). The dual model employs a random feature-based kernel mapping and an attention formulation, where $f(x)=W\phi(x)$. Training involves one step of gradient update (stochastic gradient descent, lr = 0.005) with a mean squared error loss. The detailed construction of training and test data is provided in Theorem~\ref{theorem1}.
For the simplified LLM model, the parameters $W_V$, $W_K$, and $W_Q$ are randomly initialized and remain fixed throughout use.
For evaluation, we measure the $L_2$ distance between the attention output ${\boldsymbol{h}}_{T}^{m+1}$ and the dual-model prediction $y_{\text{test}}$. The parameter $w$ in $\phi(x)$ is sampled from a standard normal distribution $\mathcal{N}(0,I)$.
We observe that as the amount of training data increases, the difference between the dual model $f(x)$ and the softmax attention layer $Transformer(x)$ outputs gradually decreases. After training the dual model for $N = 18$ steps, \textbf{this difference becomes zero}. This observation is consistent with the analysis in Theorem~\ref{theorem1}, effectively demonstrating that, \textit{under Theorem~\ref{theorem1} conditions, prompting can be equivalent to fine-tuning.}

\subsubsection{Discussion of baseline performance}
\label{Discussion of baseline performance}
The SOTA LLM-based baseline, MM-ChatAlign, suffers from two key limitations: 1) it processes all multimodal information simultaneously, relying on a scoring mechanism to balance different modalities, which makes it sensitive to weight settings; 2) this design is heavily dependent on large-scale parameters, as its original paper reports a significant 31\% performance gap between 13B and 72B LLMs on the ICWIKI dataset. In contrast, PTFEA's progressive design requires no scoring mechanism for the vast majority (over 94\%) of entities aligned in Stages I-II, thereby eliminating the sensitivity issue and heavy reliance on large models for most cases. The complex scoring is reserved only for the hardest 6\% of entities in Stage III.

\subsection{Ablation Study}
\begin{tcolorbox} [
    colback=blue!5!white,
    boxsep=0.3pt 
] 
RQ2: What roles do the different modules of PTFEA play respectively? How is the resource consumption of PTFEA?
\end{tcolorbox}
\vspace{-1em}

\subsubsection{The Importance of Progressive Inference}
We conducted an ablation study on the components of PTFEA using the ICWIKI.

\begin{figure}
\centering
\includegraphics[width=\linewidth]{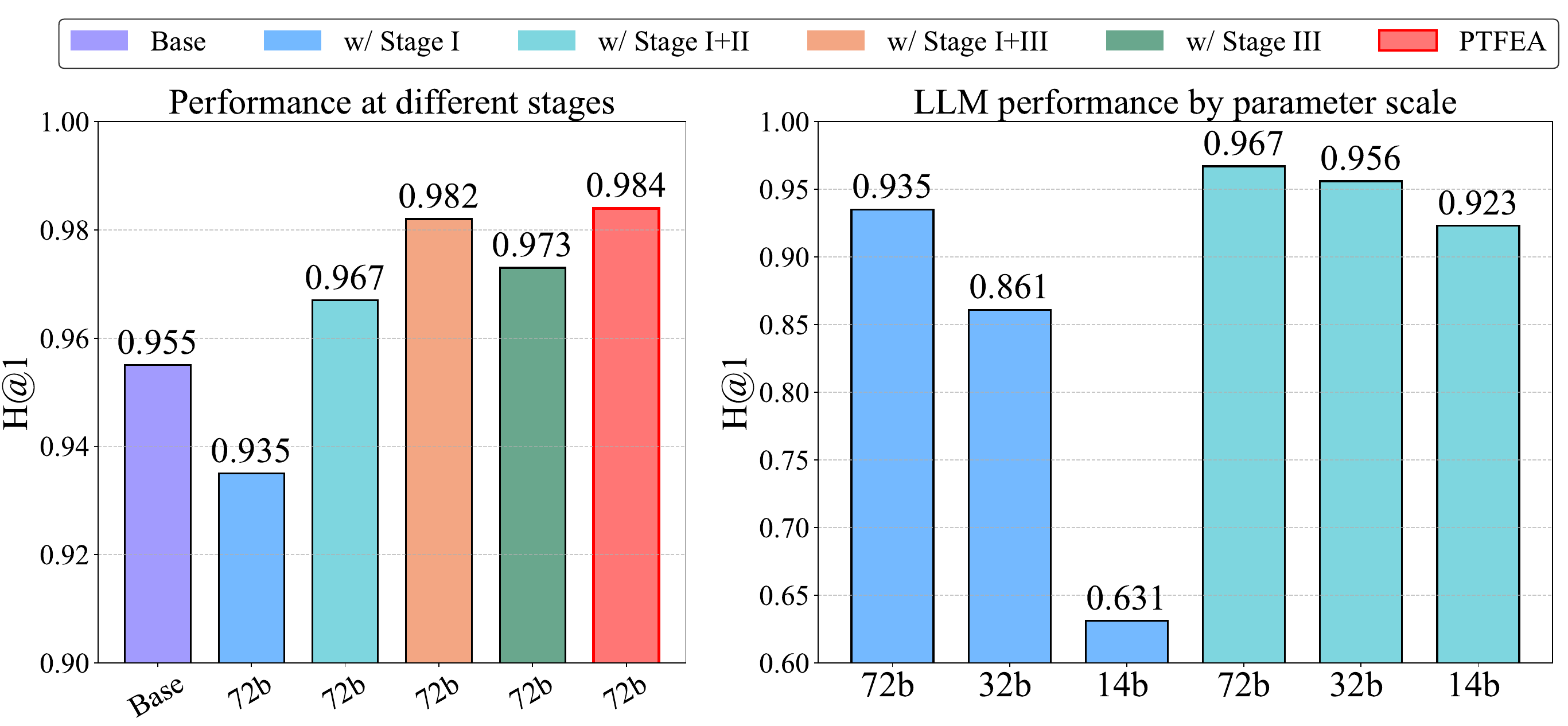}
\vspace{-2em}
\caption{
Ablation results on the ICWIKI dataset.
``Base" denotes the results of the hybrid retrieval baseline, where ``w/" indicates the inclusion of a specific component or strategy.
}
\label{ablation}
\vspace{-1em}
\end{figure}

\textbf{Effectiveness of Curriculum Learning.}
As shown in Fig.~\ref{ablation}, we conducted ablation experiments on the ICWIKI dataset. Here, \textit{``Base''} denotes embedding-based hybrid retrieval, and \textit{``w/''} means the inclusion of that module together with the Base module. The experimental results show that all three stages (Stage \uppercase\expandafter{\romannumeral 1}, ~\uppercase\expandafter{\romannumeral 2}, and \uppercase\expandafter{\romannumeral 3}) significantly impact the final performance.
For example, when Stage \uppercase\expandafter{\romannumeral 3} is removed from PTFEA, the $H@1$ score drops from 98.4\% to 96.7\%, \textbf{a decrease of 1.7\%}. Furthermore, when only Stage \uppercase\expandafter{\romannumeral 1} is retained, the $H@1$ score drops by 2\% compared to the hybrid embedding method. This is mainly because providing only the name information can lead to hallucination effects in the LLMs, reducing its confidence and overall performance, whereas embedding-based baseline methods mitigate this issue by incorporating structural, relational, and other types of information.
Furthermore, when the curriculum learning strategy is not adopted—for example, by treating all entities as hard samples (i.e., the ``w/ stage \uppercase\expandafter{\romannumeral 3}” setting shown on the left side of Fig.~\ref{ablation})—\textbf{both performance and resource efficiency decline.} Specifically, the $H@1$ accuracy decreases by 1.1\%, while resource consumption increases significantly. 

\begin{figure}
\centering
\includegraphics[width=\linewidth]{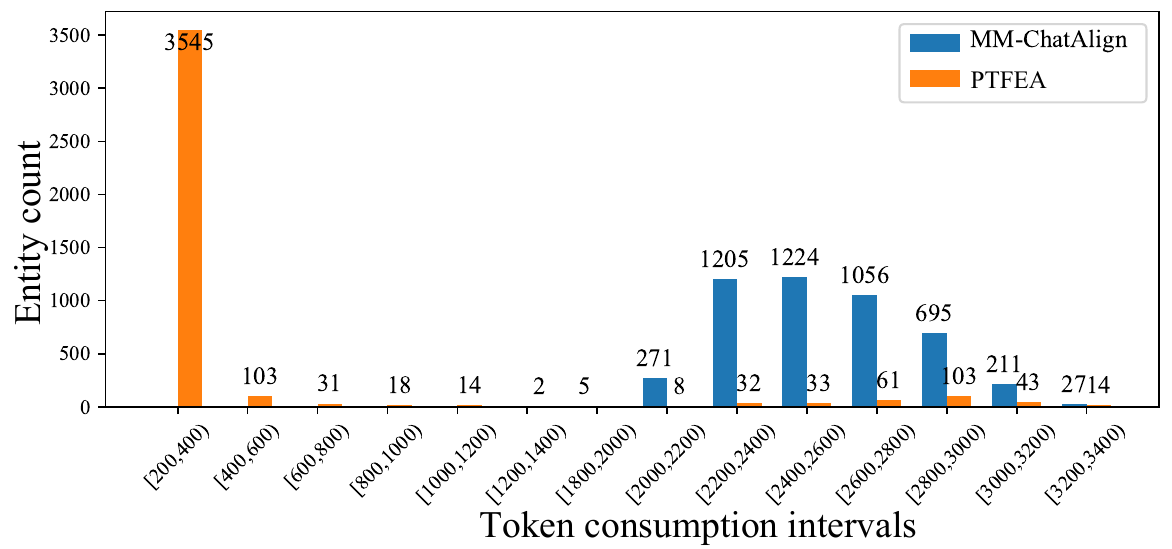}
\vspace{-2.5em}
\caption{Token consumption statistics of PTFEA and MM-ChatAlign on the ICWIKI dataset. Most of PTFEA's token usage falls within the 200–400 range.
}
\label{tokens_con}
\vspace{-2em}
\end{figure}

\textbf{Impact of Curriculum-Guided Progressive Inference for Smaller LLMs.}
As shown on the right side of Fig.~\ref{ablation}, LLMs with different parameter scales exhibit significantly different performance across stages. For example, the 72B LLM achieves an $H@1$ of 93.5\% at Stage \uppercase\expandafter{\romannumeral 1}, while the 14B LLM only reaches 63.1\%. However, the 14B LLM demonstrates a substantial improvement in Stage \uppercase\expandafter{\romannumeral 2}, reaching an $H@1$ of 92.3\%, which significantly narrows the gap compared to the 72B LLM’s 96.7\% at the same stage. Specifically, the 14B LLM achieves a relative improvement of \textbf{46.3\%,} far exceeding the 72B LLM’s 3.4\%.
\textit{These results suggest that even LLMs with smaller parameter sizes can approach the performance of larger ones when guided by curriculum learning, strongly validating the motivation for a progressive, curriculum-based approach.}

\begin{table}[htbp]
\caption{Comparison of running times under different settings.}
\small
\begin{tabular}{cccc}
\hline
Datasets & Phase & Model & Times \\ \hline
ICWIKI & PTFEA & Deepseek-v3 & 1hour, 53min, 57sec \\
ICWIKI & PTFEA & Qwen2.5-14B & 4hour, 42min, 21sec \\
ICWIKI & PTFEA & Qwen2.5-32B & 3hour, 53min, 24sec \\
ICWIKI & PTFEA & Qwen2.5-72B & 1hour, 10min, 44sec \\
ICWIKI & w/o stage III & Qwen2.5-72B & 40min, 44sec \\
ICWIKI & w/o stage II+III & Qwen2.5-72B & 39min, 17sec \\
ICWIKI & MM-ChatAlign & Qwen2.5-72B & 21hour, 11min, 21sec \\
ICYAGO & PTFEA & Deepseek-v3 & 9hour, 50min, 06sec \\ \hline
\end{tabular}
\label{times-con}
\vspace{-1em}
\end{table}

\subsubsection{Resource Consumption}
\label{Resource Consumption}
As shown in Tab.~\ref{times-con} and Fig.~\ref{tokens_con}, we evaluated the runtime and average token consumption of different methods on the ICWIKI.
The experimental results demonstrate that PTFEA significantly outperforms MM-ChatAlign in terms of runtime and token efficiency.

\textbf{Time Consumption.} Under the same configuration using the Qwen2.5-72B LLM, PTFEA requires approximately 1 hour to complete, while MM-ChatAlign takes around 21 hours, indicating that PTFEA reduces computational cost by approximately 95\% through its progressive inference mechanism. This substantial improvement is mainly attributed to the fact that, in Stage \uppercase\expandafter{\romannumeral 1} and Stage \uppercase\expandafter{\romannumeral 2}, a large amount of alignment information is effectively supplemented in advance, enabling the model to predict many alignments early and thereby avoiding the need to perform the more complex Stage \uppercase\expandafter{\romannumeral 3} full-reasoning alignment prediction for most candidate pairs.

\textbf{Token Consumption.} 
As shown in Fig.~\ref{tokens_con}, we analyze the token consumption of PTFEA and MM-ChatAlign on the ICWIKI dataset. The x-axis represents the token consumption intervals, and the y-axis indicates the number of entities falling into each interval. It can be observed that PTFEA's token usage is mainly concentrated in the 200–400 range, while MM-ChatAlign's token consumption is primarily distributed between 2200 and 3000. This demonstrates that PTFEA can significantly reduce token consumption for LLMs. The main reason is that the 72B-scale LLM already possesses strong language understanding capabilities, enabling it to generate a large number of alignment results during Stage I and II.

\subsubsection{Robustness Analysis}
\label{Robustness analysis}
To investigate the impact of missing visual modality on PTFEA, we conducted experiments on the ICWIKI dataset. The experimental results show that, compared with zero-vector imputation, employing Gaussian imputation~\cite{chen2023meaformer,mclea}, which samples from the statistics of existing embeddings, leads to a modest but consistent performance improvement. Specifically, $H@1$ increases from 98.2\% to 98.4\%, achieving a gain of 0.2\%. These results indicate that PTFEA is able to alleviate the adverse effects of missing visual modality to a certain extent.

\subsection{Effects of Different LLMs}
\begin{tcolorbox} [
    colback=blue!5!white,
    boxsep=0.3pt 
] 
RQ3: How does PTFEA perform across different LLMs?
\end{tcolorbox}
\begin{figure}
\centering
\includegraphics[width=\linewidth]{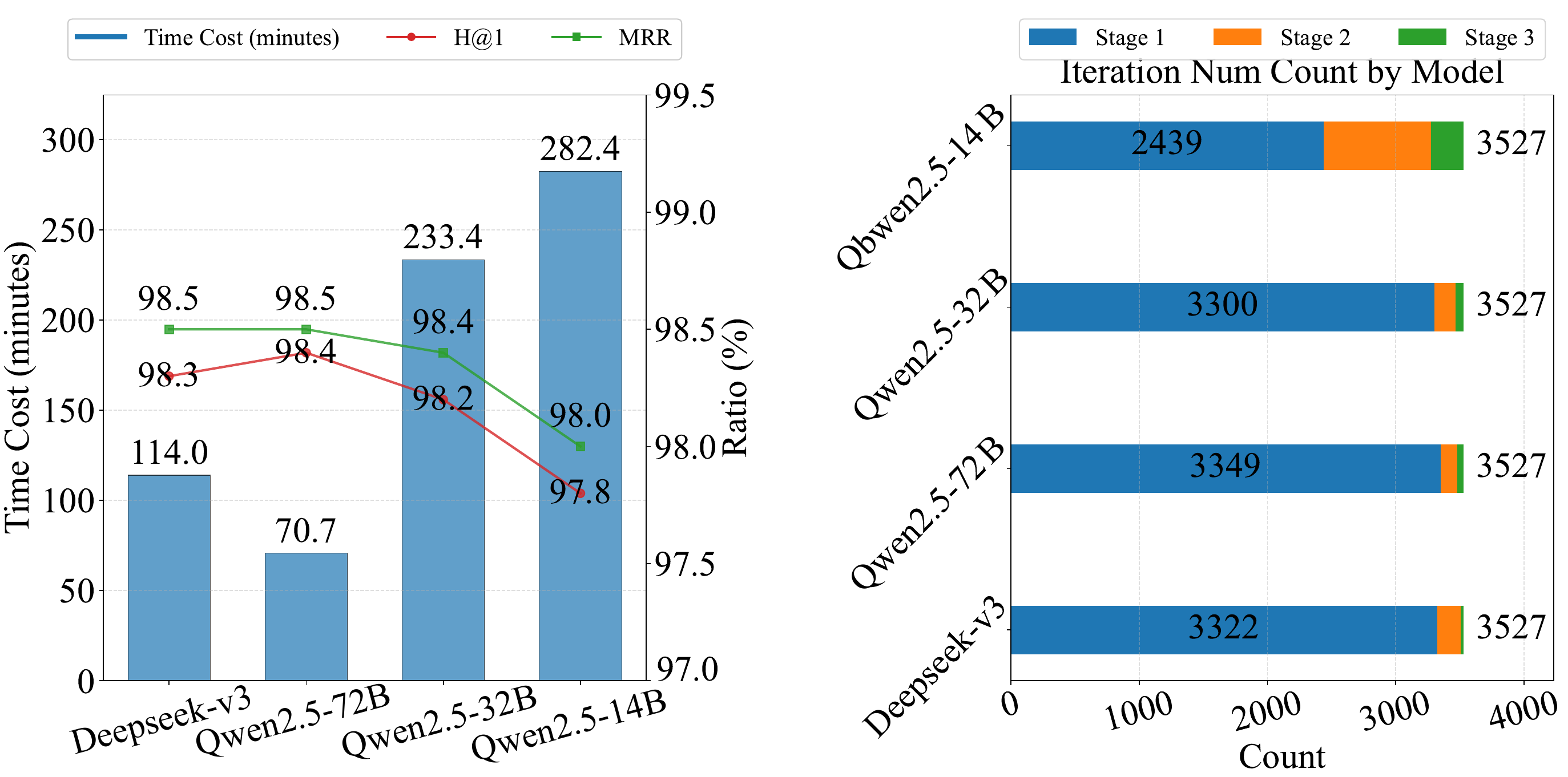}
\vspace{-1.5em}
\caption{Performance comparison of different LLMs on the ICWIKI dataset.
}
\label{diffllm}
\end{figure}

To address Research Question 3, we conducted experiments on the ICWIKI dataset. 
As shown in Fig.~\ref{diffllm}, the results indicate that \textit{the performance of PTFEA is only slightly affected by the scale of the LLMs}, with the primary differences lying in resource consumption. Overall, the best-performing model, Qwen2.5-72B, outperformed the worst-performing model, Qwen2.5-14B, by \textbf{only 0.6\% in $H@1$}, and the other LLMs also maintained $H@1$ scores around 98.2\%.
However, smaller models led to significantly longer runtimes. For example, Qwen2.5-72B completed the task in just 70.7 minutes, whereas Qwen2.5-14B took 282.4 minutes. The main reason for this discrepancy is that smaller models struggle to return a sufficient number of confident answers in Stage I. As shown in the right panel of Fig.~\ref{diffllm}, Qwen2.5-14B returned only 2,439 confident matches in Stage I, compared to 3,349 from Qwen2.5-72B—a reduction of 27.2\%.

\section{Conclusion}

In this paper, we provide the first theoretical framework unifying context engineering and model fine-tuning in the MMEA setting. Through mathematical derivation, we show that the softmax attention mechanism in LLMs is equivalent to the linear attention layer in fine-tuned models when processing multimodal entity attributes, relational context, and task descriptions. This theoretical unification enables a principled design of interpretable and efficient context engineering strategies. 
Building on this insight, we propose PTFEA, a curriculum learning-based framework that transforms traditional fine-tuning strategies into progressive prompting procedures. This approach effectively enhances the ability of LLMs of varying scales to understand and align entity information. Comprehensive experiments on five public MMEA datasets show that PTFEA consistently outperforms strong baselines. We believe this work provides a novel theoretical framework and methodological foundation for applying LLMs to MMEA tasks. In future work, we will continue to migrate other fine-tuning strategies to inspire the design of new context engineering approaches.

\section{Acknowledgements}
This work was supported by the National Natural Science Foundation of China (Grant No. 62120106008), Anhui Provincial Science and Technology Fortification Plan (Grant No. 202423k09020015), the Hefei Key Generic Technology Research and Development Program (No. 2024SGJ010), the Youth Talent Support Program of the Anhui Association for Science and Technology (Grant No. RCTJ202420), New Chongqing Youth Innovation Talent Project under Grant  CSTB2024NSCQ-QCXMX0035, and the Open Project of Key Laboratory of Knowledge Engineering with Big Data (the Ministry of Education of China), under grant number BigKEOpen2025-03. The computation was completed on the HPC Platform of Hefei University of Technology. 
 The authors would also like to acknowledge partial support from the Hefei AI Computing Center (Project Team).
 Y. He was not supported by any of these funds.
\clearpage

\bibliographystyle{ACM-Reference-Format}
\bibliography{sample-base}

@inproceedings{DBLP:conf/acl/BuCCDW0025,
  author       = {Chenyang Bu and
                  Guojie Chang and
                  Zihao Chen and
                  CunYuan Dang and
                  Zhize Wu and
                  Yi He and
                  Xindong Wu},
  title        = {Query-Driven Multimodal GraphRAG: Dynamic Local Knowledge Graph Construction
                  for Online Reasoning},
  booktitle    = {Findings of the Association for Computational Linguistics, {ACL} 2025,
                  Vienna, Austria, July 27 - August 1, 2025},
  pages        = {21360--21380},
  publisher    = {ACL},
  year         = {2025}
}

@article{zang2026medical,
  title={Medical multimodal entity linking under modality missingness},
  author={Zang, Shiji and Bu, Chenyang and Hong, Yunpeng and Ren, He and Ding, Weiping},
  journal={Artificial Intelligence in Health},
  issn={3041-0894},
  pages={026070015},
  doi={https://doi.org/10.36922/AIH026070015},
  url={https://accscience.com/journal/AIH/articles/online_first/7820},
  year={2026}
}

@inproceedings{ren2024towards,
 author = {Ren, Ruifeng and Liu, Yong},
 booktitle = {Proceedings of Advances in Neural Information Processing Systems},
 pages = {892--933},
 publisher = {Curran Associates, Inc.},
 title = {Towards Understanding How Transformers Learn In-context Through a Representation Learning Lens},
 volume = {37},
 year = {2024}
}

@article{NEURIPS2024_1b57aadd,
  title={Entity alignment with noisy annotations from large language models},
  author={Chen, Shengyuan and Zhang, Qinggang and Dong, Junnan and Hua, Wen and Li, Qing and Huang, Xiao},
  journal={Proceedings of Advances in Neural Information Processing Systems},
  volume={37},
  publisher = {Curran Associates, Inc.},
  pages={15097--15120},
  year={2024}
}

@inproceedings{lu2025breaking,
  title={Breaking the Noise Barrier: LLM-Guided Semantic Filtering and Enhancement for Multi-Modal Entity Alignment},
  author={Lu, Chenglong and Li, Chenxiao and Cheng, Jingwei and Ji, Yongquan and Chen, Guoqing and Zhang, Fu},
  booktitle={Proceedings of Conference on Empirical Methods in Natural Language Processing},
  pages={33141--33155},
publisher = {ACM},
  year={2025}
}

@inproceedings{
sahoo2025the,
title={The Diffusion Duality},
author={Subham Sekhar Sahoo and Justin Deschenaux and Aaron Gokaslan and Guanghan Wang and Justin T Chiu and Volodymyr Kuleshov},
booktitle={Proceedings of International Conference on Machine Learning},
year={2025},
url={https://openreview.net/forum?id=9P9Y8FOSOk}
}

@inproceedings{chen2025noise,
  title={Noise-powered multi-modal knowledge graph representation framework},
  author={Chen, Zhuo and Fang, Yin and Zhang, Yichi and Guo, Lingbing and Chen, Jiaoyan and Pan, Jeff Z and Chen, Huajun and Zhang, Wen},
  booktitle={Proceedings of International Conference on Computational Linguistics},
  publisher = {ACM},
  pages={141--155},
  year={2025}
}

@inproceedings{su2025mitigating,
  title={Mitigating Modality Bias in Multi-modal Entity Alignment from a Causal Perspective},
  author={Su, Taoyu and Sheng, Jiawei and Ma, Duohe and Li, Xiaodong and Yue, Juwei and Song, Mengxiao and Tang, Yingkai and Liu, Tingwen},
  booktitle={Proceedings of International ACM SIGIR Conference on Research and Development in Information Retrieval},
  pages={1186--1196},
  publisher = {ACM},
  year={2025}
}

@inproceedings{wang2025otmea,
  title={OTMEA: Multi-modal Entity Alignment via Optimal Transport},
  author={Wang, Cunda and Wang, Weihua and Li, Xinyu and Liang, Qiuyu and Bao, Feilong},
  booktitle={Proceedings of IEEE International Conference on Acoustics, Speech and Signal Processing},
  pages={1--5},
  year={2025},
  publisher={IEEE}
}

@inproceedings{li2025exploring,
  title={Exploring the Impacts of Feature Fusion Strategy in Multi-modal Entity Alignment},
  author={Li, Chenxiao and Cheng, Jingwei and Tong, Qiang and Zhang, Fu},
  booktitle={Proceedings of International Conference on Computational Linguistics},
  pages={7809--7818},
    publisher = {ACL},
  year={2025}
}

@inproceedings{feng2025cateea,
  title={CateEA: Enhancing Entity Alignment via Implicit Category Supervision},
  author={Feng, Guan Dong and Ren, Tao and Hu, Jun and dan Wang, Dan},
  booktitle={Proceedings of International Conference on Computational Linguistics},
  pages={5975--5986},
  publisher = {ACL},
  year={2025}
}

@inproceedings{agarwal2022estimating,
  title={Estimating example difficulty using variance of gradients},
  author={Agarwal, Chirag and D'souza, Daniel and Hooker, Sara},
  booktitle={Proceedings of the IEEE/CVF Conference on Computer Vision and Pattern Recognition},
  pages={10368--10378},
publisher ={IEEE},
  year={2022}
}

@inproceedings{xia2025improving,
  title={Improving complex reasoning over knowledge graph with logic-aware curriculum tuning},
  author={Xia, Tianle and Ding, Liang and Wan, Guojia and Zhan, Yibing and Du, Bo and Tao, Dacheng},
  booktitle={Proceedings of the AAAI Conference on Artificial Intelligence},
  volume={39},
  number={12},
  pages={12881--12889},
  year={2025},
publisher = {AAAI Press}
}

@inproceedings{wang2025end,
  title={From end-to-end to step-by-step: Learning to abstract via abductive reinforcement learning},
  author={Wang, Zilong and Wang, Jiongda and Chen, Xiaoyong and Wang, Meng and Ma, Ming and Wang, Zhipeng and Zhou, Zhenyu and Yang, Tianming and Dai, Wang-Zhou},
  booktitle={Proceedings of International Joint Conference on Artificial Intelligence},
  pages={6515--6523},
  year={2025},
publisher = {IJCAI Organization}
}

@article{zhang2025graph,
  title={Graph structure prefix injection transformer for multi-modal entity alignment},
  author={Zhang, Yan and Luo, Xiangyu and Hu, Jing and Zhang, Miao and Xiao, Kui and Li, Zhifei},
  journal={Information Processing \& Management},
  volume={62},
  number={3},
  pages={104048},
  year={2025},
  publisher={Elsevier}
}

@article{zhao2025me3a,
  title={ME3A: A Multimodal Entity Entailment framework for multimodal Entity Alignment},
  author={Zhao, Yu and Zhang, Ying and Sui, Xuhui and Cai, Xiangrui},
  journal={Information Processing \& Management},
  volume={62},
  number={1},
  pages={103951},
  year={2025},
  publisher={Elsevier}
}

@inproceedings{yang2024advancing,
  title={Advancing Cross-Lingual Entity Alignment with Large Language Models: Tailored Sample Segmentation and Zero-Shot Prompts},
  author={Yang, Linyan and Cheng, Jingwei and Zhang, Fu},
  booktitle={Findings of the Association for Computational Linguistics: EMNLP},
  pages={8122--8138},
publisher ={ACL},
  year={2024}
}

@misc{llmea,
      title={Two heads are better than one: Integrating knowledge from knowledge graphs and large language models for entity alignment}, 
      author={Linyao Yang and Hongyang Chen and Xiao Wang and Jing Yang and Fei-Yue Wang and Han Liu},
      year={2024},
      eprint={2401.16960},
      archivePrefix={arXiv},
      url={https://arxiv.org/abs/2401.16960}, 
}

@inproceedings{jiang2024toward,
  title={Toward practical entity alignment method design: Insights from new highly heterogeneous knowledge graph datasets},
  author={Jiang, Xuhui and Xu, Chengjin and Shen, Yinghan and Wang, Yuanzhuo and Su, Fenglong and Shi, Zhichao and Sun, Fei and Li, Zixuan and Guo, Jian and Shen, Huawei},
  booktitle={Proceedings of the ACM Web Conference},
  pages={2325--2336},
  year={2024},
  publisher={ACM}
}

@misc{chen2024llm,
      title={LLM-Align: Utilizing Large Language Models for Entity Alignment in Knowledge Graphs}, 
      author={Xuan Chen and Tong Lu and Zhichun Wang},
      year={2024},
      eprint={2412.04690},
      archivePrefix={arXiv},
      url={https://arxiv.org/abs/2412.04690}, 
}

@inproceedings{jiang2024mm,
  title={MM-ChatAlign: A novel multimodal reasoning framework based on large language models for entity alignment},
  author={Jiang, Xuhui and Shen, Yinghan and Shi, Zhichao and Xu, Chengjin and Li, Wei and Zihe, Huang and Guo, Jian and Wang, Yuanzhuo},
  booktitle={Findings of the Association for Computational Linguistics: EMNLP},
  pages={2637--2654},
publisher={ACM},
  year={2024}
}

@inproceedings{TEA,
    title = "From alignment to entailment: A Uuified textual entailment framework for entity alignment",
    author = "Zhao, Yu  and
      Wu, Yike  and
      Cai, Xiangrui  and
      Zhang, Ying  and
      Zhang, Haiwei  and
      Yuan, Xiaojie",
    booktitle = "Findings of the Association for Computational Linguistics",
    year = "2023",
    publisher = "ACL",
    pages = "8795--8806"
}

@inproceedings{nair-etal-2024-curriculum,
    title = "Curriculum learning for small code language models",
    author = {Na{\"i}r, Marwa  and
      Yamani, Kamel  and
      Lhadj, Lynda  and
      Baghdadi, Riyadh},
    booktitle = "Proceedings of the Association for Computational Linguistics",
    year = "2024",
    publisher = "ACL",
    pages = "390--401",
  
}

@inproceedings{Curriculum_learning,
author = {Bengio, Yoshua and Louradour, J\'{e}r\^{o}me and Collobert, Ronan and Weston, Jason},
title = {Curriculum learning},
year = {2009},
publisher = {ACM},
booktitle = {Proceedings of International Conference on Machine Learning},
pages = {41–48},
numpages = {8}
}

@inproceedings{wang2019dynamic,
  title={Dynamic curriculum learning for imbalanced data classification},
  author={Wang, Yiru and Gan, Weihao and Yang, Jie and Wu, Wei and Yan, Junjie},
  booktitle={Proceedings of the IEEE/CVF International Conference on Computer Vision},
  pages={5017--5026},
publisher={IEEE},
  year={2019}
}

@inproceedings{guo2020fine,
  title={Fine-tuning by curriculum learning for non-autoregressive neural machine translation},
  author={Guo, Junliang and Tan, Xu and Xu, Linli and Qin, Tao and Chen, Enhong and Liu, Tie-Yan},
  booktitle={Proceedings of the AAAI Conference on Artificial Intelligence},
  volume={34},
  number={05},
  pages={7839--7846},
publisher = "AAAI Press",
  year={2020}
}

@INPROCEEDINGS{sdea,
  author={Zhong, Ziyue and Zhang, Meihui and Fan, Ju and Dou, Chenxiao},
  booktitle={Proceedings of IEEE International Conference on Data Engineering}, 
  title={Semantics driven embedding learning for effective entity alignment}, 
  year={2022},
  volume={},
publisher={IEEE},
  number={},
  pages={2127-2140}}

@inproceedings{xgea,
  title={Cross-modal graph attention network for entity alignment},
  author={Xu, Baogui and Xu, Chengjin and Su, Bing},
  booktitle={Proceedings of ACM International Conference on Multimedia},
  pages={3715--3723},
publisher={ACM},
  year={2023}
}

@misc{cuadron2025danger,
      title={The danger of overthinking: Examining the reasoning-action dilemma in agentic tasks}, 
      author={Alejandro Cuadron and Dacheng Li and Wenjie Ma and Xingyao Wang and Yichuan Wang and Siyuan Zhuang and Shu Liu and Luis Gaspar Schroeder and Tian Xia and Huanzhi Mao and Nicholas Thumiger and Aditya Desai and Ion Stoica and Ana Klimovic and Graham Neubig and Joseph E. Gonzalez},
      year={2025},
      url={https://arxiv.org/abs/2502.08235}, 
}

@inproceedings{jiang2024unlocking,
    title = "Unlocking the Power of Large Language Models for Entity Alignment",
    author = "Jiang, Xuhui  and
      Shen, Yinghan  and
      Shi, Zhichao  and
      Xu, Chengjin  and
      Li, Wei  and
      Li, Zixuan  and
      Guo, Jian  and
      Shen, Huawei  and
      Wang, Yuanzhuo",
    booktitle = "Proceedings of  Association for Computational Linguistics",
    year = "2024",
    publisher = "ACL",
    pages = "7566--7583",
}

@inproceedings{ack-mmea,
author = {Li, Qian and Guo, Shu and Luo, Yangyifei and Ji, Cheng and Wang, Lihong and Sheng, Jiawei and Li, Jianxin},
title = {Attribute-consistent knowledge graph representation learning for multi-modal entity alignment},
year = {2023},
publisher = {ACM},
booktitle = {Proceedings of the ACM Web Conference},
pages = {2499–2508},
numpages = {10}
}

@inproceedings{AttrMMEA,
  title={Exploring and evaluating attributes, values, and structures for entity alignment},
  author={Liu, Zhiyuan and Cao, Yixin and Pan, Liangming and Li, Juanzi and Liu, Zhiyuan and Chua, Tat-Seng},
  booktitle={Proceedings of Empirical Methods in Natural Language Processing},
  year={2020},
pages="6355--6364",
publisher={ACL}
}

@article{zhu2023mmiea,
  title={MMIEA: Multi-modal interaction entity alignment model for knowledge graphs},
  author={Zhu, Bin and Wu, Meng and Hong, Yunpeng and Chen, Yi and Xie, Bo and Liu, Fei and Bu, Chenyang and Ding, Weiping},
  journal={Information Fusion},
  volume={100},
  pages={101935},
  year={2023},
  publisher={Elsevier}
}

@inproceedings{chen2020mmea,
  title={Mmea: {Entity alignment for multi-modal knowledge graph}},
  author={Chen, Liyi and Li, Zhi and Wang, Yijun and Xu, Tong and Wang, Zhefeng and Chen, Enhong},
  booktitle={Proceeding of Knowledge Science, Engineering and Management},
  pages={134--147},
  publisher={Springer},
  year={2020}
}

@inproceedings{MSNEA,
  title={Multi-modal siamese network for entity alignment},
  author={Chen, Liyi and Li, Zhi and Xu, Tong and Wu, Han and Wang, Zhefeng and Yuan, Nicholas Jing and Chen, Enhong},
  booktitle={Proceedings of SIGKDD Conference on Knowledge Discovery and Data Mining},
  pages={118--126},
  year={2022},
  publisher = {ACM},
}

@inproceedings{chen2023meaformer,
  title={Meaformer: Multi-modal entity alignment transformer for meta modality hybrid},
  author={Chen, Zhuo and Chen, Jiaoyan and Zhang, Wen and Guo, Lingbing and Fang, Yin and Huang, Yufeng and Zhang, Yichi and Geng, Yuxia and Pan, Jeff Z. and Song, Wenting and Chen, Huajun},
  booktitle={Proceedings of International Conference on Multimedia},
  pages={3317--3327},
publisher = {ACM},
  year={2023}
}

@inproceedings{liu2021visual,
  title={Visual pivoting for (unsupervised) entity alignment},
  author={Liu, Fangyu and Chen, Muhao and Roth, Dan and Collier, Nigel},
  booktitle={Proceedings of the AAAI Conference on Artificial Intelligence},
  volume={35},
  number={5},
  pages={4257--4266},
publisher = {AAAI Press},
  year={2021}
}

@inproceedings{mclea,
   author = {Zhenxi Lin and Ziheng Zhang and Meng Wang and Yinghui Shi and Xian Wu and Yefeng Zheng},
   booktitle = {Proceedings of International Conference on Computational Linguistics},
   pages = {2572-2584},
   title = {Multi-modal contrastive representation learning for entity alignment},
   year = {2022},
    publisher={ACL}
}

@inproceedings{sun2018bootstrapping,
  author = {Sun, Zequn and Hu, Wei and Zhang, Qingheng and Qu, Yuzhong},
title = {Bootstrapping entity alignment with knowledge graph embedding},
year = {2018},
publisher = {AAAI},
booktitle = {Proceedings of International Joint Conference on Artificial Intelligence},
pages = {4396–4402},
numpages = {7}
}

@inproceedings{ni2023psnea,
  title={Psnea: Pseudo-siamese network for entity alignment between multi-modal knowledge graphs},
  author={Ni, Wenxin and Xu, Qianqian and Jiang, Yangbangyan and Cao, Zongsheng and Cao, Xiaochun and Huang, Qingming},
  booktitle={Proceedings of International Conference on Multimedia},
  pages={3489--3497},
publisher = {ACM},
  year={2023}
}

@inproceedings{He_2016_CVPR,
  title={Deep residual learning for image recognition},
  author={He, Kaiming and Zhang, Xiangyu and Ren, Shaoqing and Sun, Jian},
  booktitle={Proceedings of IEEE/CVF Conference on Computer Vision and Pattern Recognition},
  pages={770--778},
publisher ={IEEE},
  year={2016}
}

@inproceedings{retrieval,
  title={Deep supervised cross-modal retrieval},
  author={Zhen, Liangli and Hu, Peng and Wang, Xu and Peng, Dezhong},
  booktitle={Proceedings of the IEEE/CVF Conference on Computer Vision and Pattern Recognition},
  pages={10394--10403},
publisher={IEEE},
  year={2019}
}

@article{ektefaie2023multimodal,
  title={Multimodal learning with graphs},
  author={Ektefaie, Yasha and Dasoulas, George and Noori, Ayush and Farhat, Maha and Zitnik, Marinka},
  journal={Nature Machine Intelligence},
  volume={5},
  number={4},
  pages={340--350},
  year={2023},
  publisher={Nature Publishing Group UK London}
}

@BOOK{test,
   author = "Donald E. Knuth",
   title = "Seminumerical Algorithms",
   volume = 2,
   series = "The Art of Computer Programming",
   publisher = "Addison-Wesley",
   address = "Reading, MA",
   edition = "2nd",
   month = "10~" # jan,
   year = "1981",
}

@inproceedings{wang2024towards,
  title={Towards semantic consistency: Dirichlet energy driven robust multi-modal entity alignment},
  author={Wang, Yuanyi and Sun, Haifeng and Wang, Jiabo and Wang, Jingyu and Tang, Wei and Qi, Qi and Sun, Shaoling and Liao, Jianxin},
  booktitle={Proceedings of IEEE International Conference on Data Engineering},
  pages={3559--3572},
  year={2024},
  organization={IEEE}
}

@article{zhou2024scmea,
  title={SCMEA: A stacked co-enhanced model for entity alignment based on multi-aspect information fusion and bidirectional contrastive learning},
  author={Zhou, Yunfeng and Zhu, Cui and Zhu, Wenjun and Li, Hongyang},
  journal={Neural Networks},
  volume={173},
  pages={106178},
  year={2024},
  publisher={Elsevier}
}

@inproceedings{mao2021boosting,
  title={Boosting the speed of entity alignment 10$\times$: Dual attention matching network with normalized hard sample mining},
  author={Mao, Xin and Wang, Wenting and Wu, Yuanbin and Lan, Man},
  booktitle={Proceedings of the ACM Web Conference},
  pages={821--832},
publisher={ACM},
  year={2021}
}

@inproceedings{dai2022can,
    title = "Why can gpt learn in-context? Language models secretly perform gradient descent as meta-optimizers",
    author = "Dai, Damai  and
      Sun, Yutao  and
      Dong, Li  and
      Hao, Yaru  and
      Ma, Shuming  and
      Sui, Zhifang  and
      Wei, Furu",
    booktitle = "Findings of the Association for Computational Linguistics",
    year = "2023",
    publisher = "ACL",
    pages = "4005--4019",
}

@ArtifactSoftware{R,
    title = {R: A Language and Environment for Statistical Computing},
    author = {{R Core Team}},
    organization = {R Foundation for Statistical Computing},
    address = {Vienna, Austria},
    year = {2019},
    url = {https://www.R-project.org/},
}

@misc{yang2023dawn,
      title={The dawn of LMMs: Preliminary explorations with GPT-4V(ision)}, 
      author={Zhengyuan Yang and Linjie Li and Kevin Lin and Jianfeng Wang and Chung-Ching Lin and Zicheng Liu and Lijuan Wang},
      year={2023},
      eprint={2309.17421},
      archivePrefix={arXiv},
      url={https://arxiv.org/abs/2309.17421}, 
}

@inproceedings{su2024ibmea,
  title={Ibmea: Exploring variational information bottleneck for multi-modal entity alignment},
  author={Su, Taoyu and Sheng, Jiawei and Wang, Shicheng and Zhang, Xinghua and Xu, Hongbo and Liu, Tingwen},
  booktitle={Proceedings of ACM International Conference on Multimedia},
  publisher= {ACM},
  pages={4436--4445},
  year={2024}
}

@inproceedings{LoginMEA,
  author       = {Taoyu Su and
                  Xinghua Zhang and
                  Jiawei Sheng and
                  Zhenyu Zhang and
                  Tingwen Liu},
  title        = {LoginMEA: Local-to-Global Interaction Network for Multi-Modal Entity
                  Alignment},
  booktitle    = {Proceedings of European Conference on Artificial Intelligence},
  volume       = {392},
  pages        = {1173--1180},
  year         = {2024}

}

@inproceedings{
zhuo2026knowledge,
title={Knowledge Reasoning Language Model: Unifying Knowledge and Language for Inductive Knowledge Graph Reasoning},
author={Xingrui Zhuo and Jiapu Wang and Gongqing Wu and Zhongyuan Wang and Jichen Zhang and Shirui Pan and Xindong Wu},
booktitle={Proceedings of International Conference on Learning Representations},
year={2026},
url={https://openreview.net/forum?id=2g8EmFwNTB}
}

@inproceedings{hong2026psqe,
  title={PSQE: A Theoretical-Practical Approach to Pseudo Seed Quality Enhancement for Unsupervised Multimodal Entity Alignment},
  author={Hong, Yunpeng and Bu, Chenyang and Zhang, Jie and He, Yi and Wu, Di and Wu, Xindong},
  booktitle={Proceedings of ACM SIGKDD Conference on Knowledge Discovery and Data Mining},
  pages={464--475},
  publisher={ACM},
  year={2026}
}

@inproceedings{chen2026dual,
  title={Dual-Branch Multi-Granularity Network with Structured Contrastive Ranking for Cross-Modal Retrieval},
  author={Chen, Zihao and Bu, Chenyang and Ji, Shengwei and Wu, Xindong},
  booktitle={Proceedings of the ACM Web Conference 2026},
  pages={1959--1970},
  publisher={ACM},
  year={2026}
}

@inproceedings{DBLP:conf/ijcai/HuangB0025,
  author       = {Manzong Huang and
                  Chenyang Bu and
                  Yi He and
                  Xindong Wu},
  title        = {How to Mitigate Information Loss in Knowledge Graphs for GraphRAG: Leveraging Triple Context Restoration and Query-Driven Feedback},
  booktitle    = {Proceedings of International Joint Conference on Artificial Intelligence},
  pages        = {8104--8112},
  year         = {2025}
}

@article{TMEA,
  title={Tackling uncertain correspondences for multi-modal entity alignment},
  author={Chen, Liyi and Sun, Ying and Zhang, Shengzhe and Ye, Yuyang and Wu, Wei and Xiong, Hui},
  journal={Proceedings of Advances in Neural Information Processing Systems},
  volume={37},
  pages={119386--119410},
  publisher = {Curran Associates, Inc.},
  year={2024}
}

@inproceedings{DBLP:conf/aaai/HuangBHZW26,
  author       = {Manzong Huang and
                  Chenyang Bu and
                  Yi He and
                  Xingrui Zhuo and
                  Xindong Wu},
  title        = {Relink: Constructing Query-Driven Evidence Graph On-the-Fly for GraphRAG},
  booktitle    = {Proceedings of the AAAI Conference on Artificial Intelligence},
  pages        = {31202--31210},
  publisher    = {{AAAI} Press},
  year         = {2026},
}

@inproceedings{DBLP:conf/www/ZhuoWWP025,
  author       = {Xingrui Zhuo and
                  Jiapu Wang and
                  Gongqing Wu and
                  Shirui Pan and
                  Xindong Wu},
  title        = {Effective Instruction Parsing Plugin for Complex Logical Query Answering
                  on Knowledge Graphs},
  booktitle    = {Proceedings of the {ACM} on Web Conference},
  pages        = {4780--4792},
  publisher    = {{ACM}},
  year         = {2025},
}

@inproceedings{DBLP:conf/ijcai/ZhuoPWW0LW025,
  author       = {Xingrui Zhuo and
                  Shirui Pan and
                  Jiapu Wang and
                  Gongqing Wu and
                  Zan Zhang and
                  Rui Li and
                  Zizhong Wei and
                  Xindong Wu},
  title        = {Progressive Prefix-Memory Tuning for Complex Logical Query Answering on Knowledge Graphs},
  booktitle    = {Proceedings of International Joint Conference on Artificial Intelligence},
  pages        = {3716--3724},
  year         = {2025}
}

@inproceedings{pmf,
  title={Progressively modality freezing for multi-modal entity alignment},
  author={Huang, Yani and Zhang, Xuefeng and Zhang, Richong and Chen, Junfan and Kim, Jaein},
  booktitle={Proceedings of Association for Computational Linguistics},
  publisher={ACM},
  pages={3477--3489},
  year={2024}
}

@inproceedings{radford2021learning,
  title={Learning transferable visual models from natural language supervision},
    author={Radford, Alec and Kim, Jong Wook and Hallacy, Chris and Ramesh, Aditya and Goh, Gabriel and Agarwal, Sandhini and Sastry, Girish and Askell, Amanda and Mishkin, Pamela and Clark, Jack and Krueger, Gretchen and Sutskever, Ilya},
  booktitle={International Conference on Machine Learning},
  pages={8748--8763},
  year={2021},
  organization={PMLR}
}

@inproceedings{cheng2025sgmea,
  title={SGMEA: Structure-guided multimodal entity alignment},
  author={Cheng, Jingwei and Guo, Mingxiao and Zhang, Fu},
  booktitle={Proceedings of International Conference on Computational Linguistics},
  pages={7851--7861},
publisher={ACL},
  year={2025}
}

\appendix


\section{Complementary Theoretical Analysis}

\subsection{Proof of Theorem 1}
\label{theproof}
\subsubsection{Forward Propagation Process of the Dual Model}
For a dual model defined as~\cite{ren2024towards}:
\[
f(x) = W\phi(x),
\]
where \(\phi(x)\) is consistent with the previously defined Softmax Attention with Kernels, we can choose \(\phi(\cdot)\) as positive random features in the following form~\cite{ren2024towards}:
\[
\phi(\boldsymbol{x}) = e^{\boldsymbol{w}^{T}\boldsymbol{x} - \|\boldsymbol{x}\|^{2}/2},
\]
which provides an unbiased approximation of the attention mechanism.
The gradient of the loss function \(\mathcal{L}\) with respect to \(W\) is given by:
\[
\frac{\partial \mathcal{L}}{\partial W} = -\frac{1}{\xi C} \sum_{j}^k \sum_{i = 1}^N \left(V_j^{(i)} \otimes \phi(K_j^{(i)})\right).
\]
where $V_{j}^{(i)} = W_V x_{j}^{i}$ and $K_{j}^{(i)} = W_K x_{j}^{i}$.
Therefore, after one step of gradient descent, the updated learnable parameter \(\widehat{W}\) in the dual model becomes:
\[
\widehat{W} = W_0 - \xi \frac{\partial \mathcal{L}}{\partial W} = W_0 + \sum_{j}^k \sum_{i = 1}^N \frac{1}{C} \left(W_V X_j^{(i)} \otimes \phi(W_K X_j^{(i)})\right),
\]
where \(W_0\) denotes the initial weight of the dual model and \(\xi\) is the learning rate.

At the testing stage, given a query input $q=W_Q x_{T}^{m+1}$, the model prediction is:
\begin{equation}
    \begin{aligned}
y_{\text{test}} &= \widehat{W} \phi(W_Q x_{T}^{m+1}) \\
&= W_0 \phi(q) + \frac{1}{C} \sum_{j}^k \sum_{i = 1}^N \left(V_j^{(i)} \otimes \phi(K_j^{(i)})\right) \phi(q),
\end{aligned}
\label{y_fx}
\end{equation}
where \(\phi(q) = \phi(W_Q X_{T}^{m+1})\).

\begin{table*}[h]
\caption{Basic statistical information of the dataset. The statistics include 5 basic datasets. These statistics encompass the number of entities, the number of relationships, the number of attributes, the number of images, and the number of aligned entity pairs.}
\begin{tabular}{ccccccc}
\hline
Dataset & KG & Entities & Relations & Attributes & Image & EA pairs \\ \hline
\multirow{2}{*}{DBP15K (ZH-EN)} & ZH (Chinese) & 19,388 & 1,701 & 8,111 & 15,912 & \multirow{2}{*}{15,000} \\
 & EN (English) & 19,572 & 1,323 & 7,173 & 14,125 &  \\ \hline
\multirow{2}{*}{DBP15K (JA-EN)} & JA (Japanese) & 19,814 & 1,299 & 5,882 & 12,739 & \multirow{2}{*}{15,000} \\
 & EN (English) & 19,780 & 1,153 & 6,066 & 13,741 &  \\ \hline
\multirow{2}{*}{DBP15K (FR-EN)} & FR (French) & 19,661 & 903 & 4,547 & 14,174 & \multirow{2}{*}{15,000} \\
 & EN (English) & 19,993 & 1,208 & 6,422 & 13,858 &  \\ \hline
\multirow{2}{*}{ICEWS-WIKI} & ICEWS & 11,047 & 272 & - & 33,141 & \multirow{2}{*}{5,058} \\
 & WIKI & 15,896 & 226 & - & 47,688 &  \\ \hline
\multirow{2}{*}{ICEWS-YAGO} & ICEWS & 26,863 & 272 & - & 80,589 & \multirow{2}{*}{18,824} \\
 & YAGO & 22,734 & 41 & - & 68,202 &  \\ \hline
\end{tabular}
\label{tab:dataset}
\end{table*}

\subsubsection{ICL Process with One Attention Layer}
When the input features $X$ contain multiple information components, denoted as:
\[
X = [X_1,X_2,...X_k,X_T]
\]
Eq.~\ref{final_equ_h} can be rewritten as:
\[
\begin{aligned}
\widehat h_{T'}^{m + 1} &= W_V X \cdot \text{softmax} \left( \frac{(W_K X)^\top W_Q x_{T}^{m + 1}}{\sqrt{d}} \right) \\
&= \frac{1}{C'} W_V [X_1,X_2,...X_k, X_T] [\phi(W_K X_{1}), \\
& \phi(W_K X_{2}), \ldots, \phi(W_K X_T)]^\top \phi(W_Q X_{T}^{m + 1})
\end{aligned}
\]
where $C'$ is a normalization constant for the equivalent attention block. By defining $V = W_V X$ and $K = W_K X$, we derive:
\begin{equation}
    \begin{aligned}
\widehat h_{T'}^{m + 1} &= \frac{1}{C'} [V_{1}, V_{2}, \ldots, V_T] [\phi(K_{2}), \phi(K_{2}), \ldots, \phi(K_T)] \phi(q) \\
&= \frac{1}{C'} \left( V_{1} \phi(K_{1})^\top + V_{2} \phi(K_{2})^\top + \cdots + V_T \phi(K_T)^\top \right) \phi(q) \\
&= W_0' \phi(q) + \frac{1}{C'} \sum_{j}^k \sum_{i=1}^N \left( V_j^{(i)} \otimes \phi(K_j^{(i)}) \right) \phi(q)
\end{aligned}
\label{h_icl}
\end{equation}
where $W_0' = \frac{1}{C'} V_T \phi(K_T)^\top$ is the baseline weight term, and $\otimes$ denotes the outer product operation.

By comparing Eq.~\ref{y_fx} and Eq.~\ref{h_icl}, it can be observed that when \( W_0 = W_0' \) and \( C = C' \), the output of the dual model \( f(x) \) is identical to the output of an attention layer.

\subsubsection{More Discussion on Loss}
Noticing that $\mathcal{L}$ is a scalar function, it can be inferred from Eq.~\ref{h_icl} that the differential of $\mathcal{L}$ should be:
\begin{equation}
    \begin{array}{l}
d{\mathcal L} = tr( - \frac{1}{{\xi C}} \sum\limits_j^k  {\sum\limits_{i = 1}^N {\phi ({W_K}X_j^i} } ){({W_V}X_j^i)^T}dW)\\
 = tr( - \frac{1}{{\xi C}}\sum\limits_j^k  {\sum\limits_{i = 1}^N {{{({W_V}X_j^i)}^T}dW\phi ({W_K}X_j^i))} } \\
 = d(tr( - \frac{1}{{\xi C}}\sum\limits_j^k  {\sum\limits_{i = 1}^N {{{({W_V}X_j^i)}^T}W\phi ({W_K}X_j^i)))} } 
\end{array}
\end{equation}

Therefore, the loss $\mathcal{L}$ can be obtained as:
\[{\mathcal L} =  - \frac{1}{{\xi C}}\sum\limits_{j \in X} {\sum\limits_{i = 1}^N {{{({W_V}X_j^i)}^T}W\phi ({W_K}X_j^i)} } \]

\section{Supplementary Notes on Experiments}
\label{Supplementary Notes on Experiments}

\textbf{Dataset Statistics}
\label{dataset}
Our experimental datasets mainly include DBP15K (ZH-EN, JA-EN, FR-EN), as well as ICEWS-WIKI and ICEWS-YAGO. The experimental data for the first three datasets are consistent with those used in MEAformer, MCLEA, and other methods, while the data for the latter two datasets align with those in MM-ChatAlign. Our detailed dataset statistics are presented in Tab.~\ref{tab:dataset}.

\textbf{Evaluation Metrics.}
\label{Evaluation Metrics}
The most commonly used evaluation metrics for entity alignment are 
$Hits@n$ ($H@n$) and Mean Reciprocal Rank ($MRR$). Higher values of 
$Hits@n$ and $MRR$ indicate better alignment performance.
Given a test alignment set 
$S=\{(e_s^i, e_t^i)\}_{i=1}^{|S|}$, for each source entity $e_s^i$, 
the model ranks all candidate target entities according to their alignment 
scores. Let $rank_i$ denote the rank position of the ground-truth target 
entity $e_t^i$ in the ranked candidate list. $Hits@n$ measures the proportion 
of test entities whose correct counterpart is ranked within the top $n$ 
positions, and is defined as:
\begin{equation}
Hits@n = \frac{1}{|S|} \sum_{i=1}^{|S|} \mathbb{I}(rank_i \le n),
\end{equation}
where $\mathbb{I}(\cdot)$ is the indicator function, which equals 1 if the 
condition holds and 0 otherwise.
$MRR$ measures the average reciprocal rank of the correct target entities, 
and is calculated as:
\begin{equation}
\begin{aligned}
MRR &= \frac{1}{|S|} \sum_{i=1}^{|S|} \frac{1}{rank_i} \\
    &= \frac{1}{|S|} \left( \frac{1}{rank_1} + \cdots + \frac{1}{rank_{|S|}} \right).
\end{aligned}
\end{equation}

\textbf{More Ablation Study.}
More ablation experiments on different datasets are reported in Tab.~\ref{tab:ablation_datasets}.
\begin{table}[htbp]
\centering
\caption{Ablation study on different datasets ($H@1$).}
\label{tab:ablation_datasets}
\begin{tabular}{lccc}
\toprule
\textbf{Dataset} & \textbf{Base} & \textbf{Stage I} & \textbf{Stage I + II} \\
\midrule
ICYAGO  & 0.964 & 0.902 & 0.966 \\
ZH--EN  & 0.931 & 0.885 & 0.968 \\
JA--EN  & 0.965 & 0.865 & 0.968 \\
FR--EN  & 0.985 & 0.909 & 0.979 \\
\bottomrule
\end{tabular}
\vspace{-1em}
\end{table}

\begin{figure}
\centering
\includegraphics[width=\linewidth]{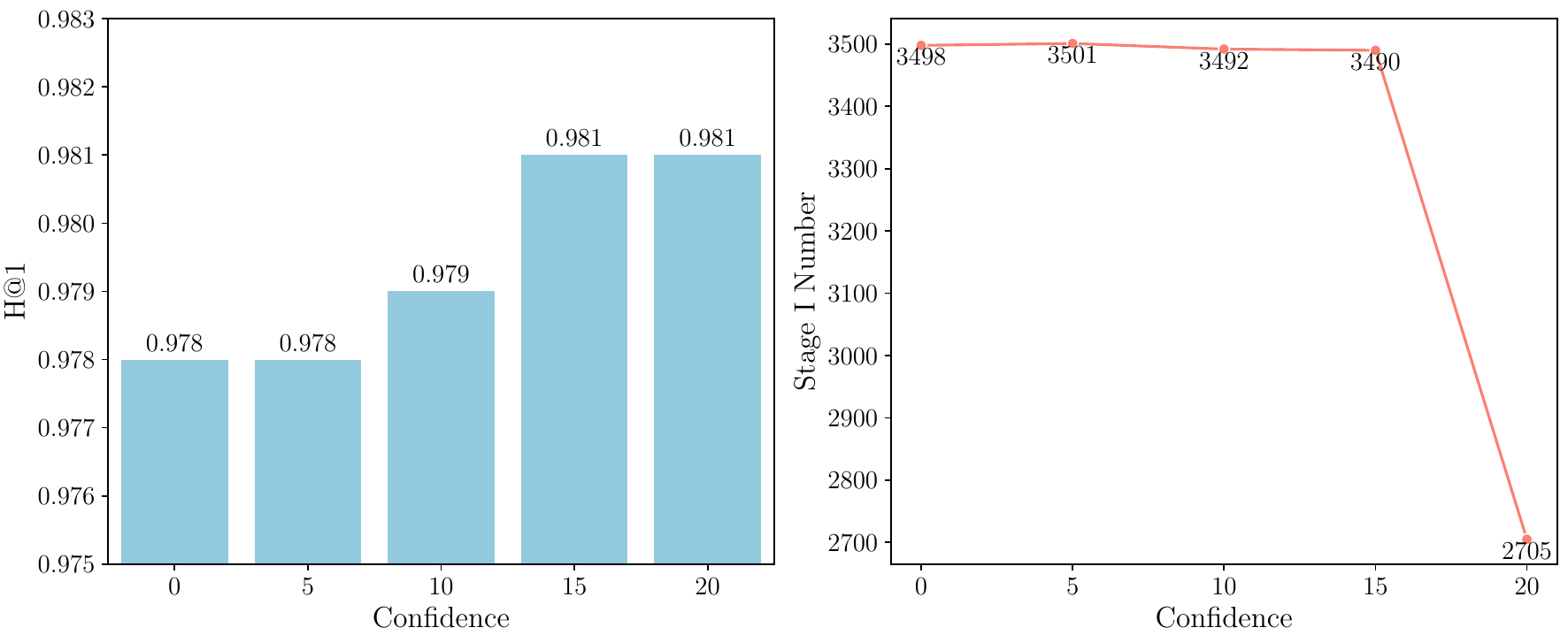}
\caption{Comparison results of fixed confidence thresholds on the ICWIKI.
The left chart shows the impact of different confidence thresholds on $H@1$, while the right chart shows their effect on the output quantity of Stage  \uppercase\expandafter{\romannumeral 1}.
}
\label{parameter}
\vspace{-1em}
\end{figure}
\textbf{Impact of Confidence Threshold.} Fig.~\ref{parameter} illustrates the impact of different confidence thresholds on the final $H@1$ performance, as well as the number of confident matches returned in Stage I. 
We observe that a lower confidence threshold (e.g., 5) results in slightly lower $H@1$ performance compared to a higher threshold (e.g., 15), with a difference of about 0.3\%. 

\end{sloppypar} 

\end{document}